\documentclass[11pt]{article}

\usepackage[utf8]{inputenc}
\usepackage{a4wide}
\usepackage{graphicx}
\usepackage{amsmath}
\usepackage{amssymb}
\usepackage{slashed}
\usepackage{listings}
\usepackage{color}
\usepackage{authblk}

\usepackage{cite}
\usepackage{fancyheadings}
\usepackage[title,titletoc]{appendix}

\numberwithin{equation}{section}

\def\preprintn{ZU-TH 24/19}

\fancypagestyle{first}
{
  
  \fancyhead[R]{\footnotesize \preprintn}
}

\usepackage[colorlinks=true
,urlcolor=blue
,anchorcolor=blue
,citecolor=blue
,filecolor=blue
,linkcolor=blue
,menucolor=blue
,pagecolor=blue
,linktocpage=true
,pdfproducer=medialab
,pdfa=true
]{hyperref}

\newcommand{\comment}[2]{#2}

\def\N{\mathcal{N}}
\def\O{\mathcal{O}}
\def\s{\mathbf{s}}

\def\z{\mathbf{z}}

\def\N{\mathcal{N}}

\def\F{\mathbb{F}}
\def\Z{\mathbb{Z}}
\def\Q{\mathbb{Q}}

\newcommand{\mmod}[1]{\ \text{mod}\ #1}

\newcommand{\mybox}[1]{\text{\fboxsep=1.0em\fbox{$\displaystyle#1$}}}

\def\la{\langle}
\def\ra{\rangle}

\def\nn{\nonumber \\}

\title{\textbf{\textsc{FiniteFlow}: multivariate functional reconstruction using
    finite fields and dataflow graphs}}

\date{}

\comment{
} 

\author[]{Tiziano Peraro}
\affil[]{\emph{\small Physik-Institut, Universit\"at Z\"urich, Wintherturerstrasse 190, CH-8057 Z\"urich, Switzerland}}

\begin{document}

\maketitle

\thispagestyle{first}

\begin{abstract}
  Complex algebraic calculations can be performed by reconstructing
  analytic results from numerical evaluations over finite fields.  We
  describe \textsc{FiniteFlow}, a framework for defining and executing
  numerical algorithms over finite fields and reconstructing
  multivariate rational functions.  The framework employs
  computational graphs, known as dataflow graphs, to combine basic
  building blocks into complex algorithms.  This allows to easily
  implement a wide range of methods over finite fields in high-level
  languages and computer algebra systems, without being concerned with
  the low-level details of the numerical implementation.  This
  approach sidesteps the appearance of large intermediate expressions
  and can be massively parallelized.  We present applications to the
  calculation of multi-loop scattering amplitudes, including the
  reduction via integration-by-parts identities to master integrals or
  special functions, the computation of differential equations for
  Feynman integrals, multi-loop integrand reduction, the decomposition
  of amplitudes into form factors, and the derivation of integrable
  symbols from a known alphabet.  We also release a proof-of-concept
  \textsc{C++} implementation of this framework, with a high-level
  interface in \textsc{Mathematica}.
\end{abstract}

\clearpage

\tableofcontents

\section{Introduction}
\label{sec:introduction}

Scientific theoretical predictions often rely on complex algebraic
calculations.  This is especially true in high energy physics, where
current and future experiments demand precise predictions for complex
scattering processes.  One key ingredient for making these predictions
are scattering amplitudes in perturbative quantum field theory.  The
complexity of these predictions depends on several factors, most
notably the loop order, where higher loop orders are required for
higher precision, the number of scattering particles involved, and the
number of independent physical scales describing the process.

A major bottleneck in many analytic predictions is the appearance of
large expressions in intermediate stages of the calculation.  These
can be orders of magnitude more complicated than the final result.
Large analytic cancellations often happen in the very last stages of a
calculation.  While computer algebra extraordinarily enhances our
capability of making such predictions, due to the reasons above, it
needs to complemented with more effective techniques when dealing with
the most challenging computations.

One can trivially observe that the mentioned bottleneck is not present
in numerical calculations with fixed precision, where every
intermediate result is a number (or a list of numbers).  However, in
some fields, high-energy physics being one of them, analytic
calculations provide more valuable results -- since they can provide a
more accurate numerical evaluation, and the possibility of further
checks, studies and manipulations-- and in some cases our only
reliable way of obtaining them.

An effective method for sidestepping the bottleneck of complex
intermediate expressions consists of reconstructing analytic
expressions from numerical evaluations.  This can be effectively used
in combination with \emph{finite fields}, i.e.\ numerical fields with
a finite number of elements.  In particular, we may choose fields
whose elements can be represented by machine size integers, where
basic operations can be done via modular arithmetic.  Numerical
operations over these fields are therefore relatively fast, but also
exact, while they avoid the need of using multi-precision arithmetic,
which is computationally expensive.  Full analytic expressions for
multivariate rational functions can then be obtained, using
\emph{functional reconstruction} techniques, from several numerical
evaluations with different input values and, if needed, over several
finite fields.  Thanks to these algorithms, the problem of computing a
rational function is reduced to the problem of providing an efficient
numerical evaluation of it over finite fields.  This implies that they
can be applied to a very broad range of problems.  Moreover, numerical
evaluations can be massively parallelized, taking full advantage of
the available computing resources.

Finite fields have been used by computer algebra systems for a long
time.  In high-energy physics, they were introduced in
ref.~\cite{vonManteuffel:2014ixa} for the solution of (univariate)
integration-by-parts~(IBP) identities.  In ref.~\cite{Peraro:2016wsq}
we developed a multivariate reconstruction algorithm which is suitable
for complex multi-scale problems and showed how to apply it to other
techniques in high-energy physics, such as integrand
reduction~\cite{Ossola:2006us,Giele:2008ve,Mastrolia:2011pr,Badger:2012dp,Zhang:2012ce,Mastrolia:2012an}
and generalized
unitarity~\cite{Bern:1994zx,Bern:1994cg,Britto:2004nc,Ellis:2007br}.
Since then, functional reconstruction techniques based on evaluations
over finite fields have been successfully employed in several
calculations which proved to be beyond the reach of conventional
computer algebra systems, with the available computing resources (see
e.g.\
ref.s~\cite{vonManteuffel:2016xki,Badger:2018enw,Abreu:2018zmy,Lee:2019zop,Henn:2019rmi,vonManteuffel:2019wbj,Abreu:2019odu,vonManteuffel:2019gpr,Badger:2019djh}
for some notable examples).

Despite the remarkable results which have already been obtained with
functional reconstruction techniques, there are still some obstacles
which prevent a more widespread usage of them.  A first one is the
lack of a public implementation of functional reconstruction
techniques suitable for arbitrary multivariate rational
functions.\footnote{During advanced stages of preparation of this
  work, an implementation of a sparse multivariate reconstruction
  algorithm was published~\cite{Klappert:2019emp}.  In this paper we
  describe instead a dense reconstruction algorithm.}  A second
obstacle is the need of providing an efficient numerical
implementation of the functions to be reconstructed, which is
typically best done in statically compiled low-level languages, such
as \textsc{C}, \textsc{C++} or \textsc{Fortran}.  In this paper, we
try to address both these problems.

Let us assume a functional reconstruction algorithm is available, and
consider the problem of providing an efficient numerical evaluation of
an algorithm representing a rational function over finite fields.  The
first possibility is obviously low-level coding.  This offers great
performance and flexibility, but it is also hard and time-consuming to
program and therefore it limits the usability of these techniques,
especially if compared with the ease of use of computer algebra
systems.

Another strategy consists in coding up some algorithms in low-level
languages and providing interfaces in higher level languages and
computer algebra systems.  This combines the efficiency of low-level
languages with the ease of use of high-level ones.  As an example,
consider the problem of solving a linear system of equations with
parametric rational entries.  Most computer algebra systems have
dedicated built-in procedures for this.  One could build another
procedure, with a similar interface, which instead sends the system to
a \textsc{C/C++} code, which in turn solves it numerically several
times and reconstructs the analytic solution from these numerical
evaluations.  For the user of the procedure, there is very little
difference (except for performance) with respect to using the built-in
procedure.  Unfortunately, this strategy strongly limits the
flexibility of functional reconstruction, since one is limited to use
a set of hardcoded algorithms.  Moreover, these algorithms often solve
only an intermediate step of a more complex calculation needed by a
scientific prediction.  For instance, in most cases, one needs to
substitute the solution of a linear system into another expression and
then perform other operations or substitutions before obtaining the
final result.  Significant analytic simplifications often occur at the
very last steps of the calculation, making thus the reconstruction of
the intermediate steps a highly inefficient strategy.  We thus need
something which is much more flexible and applicable to a wider
variety of problems.

One can observe that many different complex calculations share common
building blocks.  For instance, many calculations involve the solution
of one or more linear systems, changes of variables, linear
substitutions, and so on, in intermediate stages.  These intermediate
calculations, however, need to be combined in very different ways,
depending on the specific problem an algorithm is meant to solve.
Building on this observation, we propose a strategy which allows to
easily combine these basic building blocks into arbitrarily complex
calculations.

In this paper, we introduce a framework, that we call \textsc{FiniteFlow},
which allows to easily define complicated numerical algorithms over finite
fields, and reconstruct analytic expressions out of numerical
evaluations.  The framework consists of three main components.  The
first component is a set of basic numerical algorithms, efficiently
implemented over finite fields in a low-level language.  These include
algorithms for solving linear systems, linear fits, evaluating
polynomials and rational functions, and many more.  The second
component is a system for combining these basic algorithms, used as
building blocks, into arbitrarily more complex ones.  This is done
using \emph{dataflow graphs}, which provide a graphical representation
of a complex calculation.  Each \emph{node} in the graph represents a
basic algorithm.  The inputs of each of these algorithms are in turn
chosen to be the outputs of other basic algorithms, represented by
other nodes.  This provides a simple and effective way of defining
complicated algebraic calculations, by combining basic building blocks
into complex algorithms, without the need of any low-level coding.
Indeed, this framework can be more easily used from interfaces in
high-level languages and computer algebra systems.  Dataflow graphs
can be numerically evaluated and their output represents -- in our
framework -- a list of rational functions.  Numerical evaluations with
different inputs can be easily performed in parallel, in a highly
automated way.  Indeed, this defines an algorithm-independent strategy
for exploiting computing resources consisting of several cores, nodes,
or machines.  The third and last component consists of functional
reconstruction algorithms, which are used to reconstruct analytic
formulas out of the numerical evaluations (which in turn, as stated,
may be represented by a graph).  We propose here an improved version
of the reconstruction algorithms already presented
in~\cite{Peraro:2016wsq}.

The idea of using dataflow graphs for defining a numerical calculation
is not new.  For instance, they are notably used in the popular
\textsc{TensorFlow} library~\cite{tensorflow2015-whitepaper}, in the
context of machine learning and neural networks.  Although in this
paper, we are interested in a very different application, one can
point out a few similarities.  For instance, the \textsc{TensorFlow}
library allows to define complex functions (which, in that case, often
represent neural networks) from high-level languages, which then need
to be efficiently evaluated several times.  To the best of our
knowledge, this paper describes for the first time an application of
dataflow graphs for the purpose of defining (rational) numerical
algorithms over finite fields, to be used in combination with
functional reconstruction techniques.  In particular, we will show, by
providing several examples, that they are suited for solving many
types of important problems in high-energy physics.

With this paper, we also release a proof-of-concept \textsc{C++}
implementation of this framework, which includes a
\textsc{Mathematica} interface.  This code has already been used in a
number of complex analytic calculations, including some recently
published cutting-edge scientific
results~\cite{Badger:2018enw,Henn:2019rmi,Badger:2019djh}, and we
thus think its publication can be highly beneficial.  We stress
that \textsc{FiniteFlow} is not meant to provide the solution of any
specific scientific problem, but rather a framework which can be used
for solving a wide variety of problems.  We also provide public codes
with several packages and examples of applications of
\textsc{FiniteFlow} to very common problems in high-energy physics,
which can be easily adapted to similar problems as well.

The paper is organized as follows.  In
section~\ref{sec:finite-fields-funct} we review some basic concepts
about finite fields and rational functions, and we describe an
efficient functional reconstruction algorithm for multivariate
functions.  In section~\ref{sec:dataflow-graphs} we describe our
system for defining numerical algorithms, based on dataflow graphs.
In section~\ref{sec:numer-algor-over} we describe the implementation
of several numerical algorithms over finite fields, which are the
basic building blocks of the dataflow graphs representing a more
complex computation.  In the next sections, we describe the application
of this framework to several problems in high-energy physics.  In
section~\ref{sec:reduct-scatt-ampl} we discuss the reduction of
scattering amplitudes to master integrals or special functions, as
well as the Laurent expansion in the dimensional regulator.  In
section~\ref{sec:diff-equat-mast} we discuss the application to
differential equations for computing master integrals.  In
sections~\ref{sec:integrand-reduction}
and~\ref{sec:tens-reduct-decomp} we discuss multi-loop integrand
reduction and the decomposition of amplitudes into form factors
respectively.  In section~\ref{sec:find-integr-symb} we talk about the
derivation of integrable symbols from a known alphabet.  Finally, in
section~\ref{sec:proof-conc-impl} we give some details about our
public proof-of-concept implementation, and in
section~\ref{sec:conclusions} we draw our conclusions.

\section{Finite fields and functional reconstruction}
\label{sec:finite-fields-funct}

In this section, we set some notation by reviewing well-known facts
about finite fields and rational functions.  We also describe a
multivariate reconstruction algorithm based on numerical evaluations
over finite fields.  The latter is based on the one described
in~\cite{Peraro:2016wsq} with a few modifications and improvements.  A
slightly more thorough treatment of the subject, which uses a notation
compatible with the one of this paper, can be found in
ref.~\cite{Peraro:2016wsq} (in particular, in sections~2, 3 and
Appendix~A of that reference).

\subsection{Finite fields and rational functions}
\label{sec:finite-fields}

Finite fields are mathematical fields with a finite number of
elements.  In this paper, we are only concerned with the simplest and
most common type of finite field, namely the set of integers modulo a
prime $p$, henceforth indicated with $\Z_p$.  In general, for any
positive integer $n$, we call $\Z_n$ the set of non-negative integers
smaller than $n$.  All basic rational operations in $\Z_n$, except
division, can be trivially defined using modular arithmetic.  One can
also show that if $a\in\Z_n$ and $\textrm{gcd}(a,n)=1$ then $a$ has a
unique inverse in $\Z_n$.  In particular, if $n=p$ is prime, an inverse
exists for any non-vanishing element of $\Z_p$, hence any rational
operation is well defined.  This also defines a map between rational
numbers $q=a/b\in\Q$ and $\Z_n$, for any rational whose denominator
$b$ is coprime with $n$.  It also implies that any numerical
algorithm which consists of a sequence of rational operations can be
implemented over finite fields $\Z_p$.  In particular, polynomials
and rational functions are well defined mathematical objects.

Given a set of variables $\z = \{z_1,\ldots,z_n\}$ and a numerical
field $\F$, one can define polynomial and rational  functions of $\z$
over $\F$.  More in detail, any list of exponents
$\alpha=\{\alpha_1,\ldots,\alpha_n\}$, defines the monomial
\begin{equation}
  \label{eq:1}
  \z^\alpha = \prod_{j=1}^n\, z_j^{\alpha_j}.
\end{equation}
Polynomials over $\F$ have a unique representation as linear
combinations of monomials
\begin{equation}
  \label{eq:poly}
  p(\z) = \sum_\alpha\, c_\alpha\, \z^\alpha,
\end{equation}
with coefficients $ c_\alpha \in \F$.  Rational functions are ratios
of two polynomials
\begin{equation}
  \label{eq:ratfun}
  f(\z) = \frac{\sum_\alpha\, n_\alpha\, \z^\alpha}{\sum_\alpha\, d_\alpha\, \z^\alpha},
\end{equation}
with $n_\alpha,d_\alpha \in \F$.  Notice that the representation of
$f(\z)$ in Eq.~\eqref{eq:ratfun} is not unique.  A unique
representation can, however, be obtained by requiring numerator and
denominator to have no common polynomial factor, and fixing a
convention for the normalization on the coefficients
$n_\alpha,d_\alpha$.  We find that a useful convention is setting
$d_{\textrm{min}(\alpha)}=1$, where $\z^{\textrm{min}(\alpha)}$ is the
smallest monomial appearing in the denominator with respect to a
chosen monomial order.  Using this convention, the constant term in
the denominator, if present, is always equal to one.

An important result in modular arithmetic is Wang's \emph{rational
  reconstruction}
algorithm~\cite{Wang:1981:PAU:800206.806398,Wang:1982:PRR:1089292.1089293}
which allows, in some cases, to invert the map between $\Q$ and
$\Z_n$.  More in detail, given the image $z\in\Z_n$ of a number
$q=a/b\in \Q$, Wang's algorithm successfully reconstructs $q$ if $n$
is large enough with respect to the numerator and the denominator of
the rational number -- more precisely if and only if
$|a|, |b|<\sqrt{n/2}$.  Hence, if a prime $p$ is sufficiently large,
one can successfully reconstruct a rational number from its image in
$\Z_p$.  However, our main reason for using finite fields is the
possibility of performing calculations efficiently using machine size
integers, which on most modern machines can have a size of 64 bits.
This requirement forces us to use primes such that $p<2^{64}$.  One
can overcome this limitation by means of the Chinese remainder
theorem, which allows to deduce a number $a\in \Z_{n}$ from its images
$a_i\in \Z_{n_i}$ if the integers $n_i$ have no common factors.
Hence, given a sequence of primes $\{p_1,p_2,\ldots\}$, from the image
of a rational number over several prime fields
$\Z_{p_1},\Z_{p_2},\ldots$ one can deduce the image of the same number
over $\Z_{p_1\, p_2\dots}$.  Once the product of the selected primes
is large enough, Wang's reconstruction algorithm will be successful.

The functional reconstruction algorithm we will describe in the next
section can be performed over any field, but in practice, it will only
be implemented over finite fields.  The coefficients of the
reconstructed function (i.e.\ $n_\alpha,d_\alpha$ appearing in
Eq.~\eqref{eq:ratfun}) are then mapped over the rational field using
Wang's algorithm and checked numerically against evaluations of the
function
over other finite fields.  If the check is unsuccessful, we
proceed with reconstructing the function over more finite fields
$\Z_{p_i}$, and combine them using the Chinese remainder theorem as
explained above, in order to obtain a new result over $\Q$.  The
algorithm terminates when the result over $\Q$ agrees with numerical
checks over finite fields which have not been used for the
reconstruction.

\subsection{Multivariate functional reconstruction}
\label{sec:mult-funct-reconstr}

We now turn to the, so called, \emph{black box interpolation problem},
i.e.\ the problem of inferring, with very high probability, the
analytic expression of a function from its numerical evaluations.  We
assume to have a numerical procedure for evaluating an $n$-variate
rational function $f$, whose analytic form is not known.  More in
detail, the procedure takes as input numerical values for $\z$ and a
prime $p$ and returns the function evaluated at $\z$ over the finite
field $\Z_p$,
\begin{equation} \label{eq:blackboxmodp}
  (\z,p) \longrightarrow \mybox{f} \longrightarrow f(\z) \mmod{p}.
\end{equation}
We also allow the possibility for this procedure to fail the
evaluation.  We call this evaluation points \emph{bad points} or
\emph{singular points}.  Notice that these do not necessarily
correspond to a singularity in the analytic expression of the
function, but also to spurious singularities in intermediate steps
of the procedure, or to any other interference with the possibility of
evaluating the function with the implemented numerical algorithm.
When this happens, the singular evaluation point is simply replaced
with a different one.  We stress, however, that the occurrence of such
cases is extremely unlikely for a realistic problem, provided that the
evaluation points are chosen with care (we will expand on this later).

A functional reconstruction algorithm aims to identify the monomials
appearing in the analytic expression of the function as in
Eq.~\eqref{eq:ratfun}, and the value of their coefficients
$n_\alpha, d_\alpha$.  The basic reconstruction algorithm we discuss
in this section is based on a strategy already proposed in
ref.~\cite{Peraro:2016wsq}.  However, we find it is useful to briefly
summarize it here in order to point out a few modifications and
improvements, and also because the discussion below will benefit from
having a rough knowledge of how the functional reconstruction works.

For univariate polynomials, our reconstruction strategy is based on
Newton's polynomial representation~\cite{abramowitz1964handbook}
\begin{align} \label{eq:unewtonrep}
  f(z) ={}& \sum_{r=0}^R a_r \prod_{i=0}^{r-1} (z-y_i) \nn
        = {}& a_0 + (z-y_0)\bigg(a_1 + (z-y_1)\Big( a_2 + (z-y_2) \big(\cdots+ (z-y_{R-1})\, a_R\big) \Big) \bigg),
\end{align}
where $R$ is the total degree, and $y_0,y_1,y_2,\ldots$ are a sequence
of distinct numbers.  One can easily check that, with this
representation, any coefficient $a_r$ can be determined from the
knowledge of the value of the function at $z=y_r$ and from the
coefficients $a_j$ with $j<r$.  In particular, it does not require the
knowledge of the total degree $R$.  This allows to recursively
reconstruct the coefficients $a_r$ of the polynomial, starting from
$a_0$ which is determined by $f(y_0)$.  If the total degree of the
polynomial is not known, the termination criterion of the
reconstruction algorithm is the agreement between new evaluations of
the function $f$ and the polynomial defined by the coefficients
reconstructed so far.  In some cases, the total degree, or an upper
bound to it, is known a priori (see e.g.\ when this is used in the
context of a multivariate reconstruction) and therefore one can
terminate the reconstruction as soon as this bound is reached.  After
the polynomial is reconstructed, it is converted back into a canonical
representation.

For univariate rational functions, we distinguish two cases.  The first
case, which will be useful in the context of multivariate
reconstruction, is when the total degree of the numerator and the
denominator of the function are known and the constant term in the
denominator does not vanish.  This means, remembering the
normalization convention we introduced in
section~\ref{sec:finite-fields}, that we can parametrize the function
as
\begin{equation}
  \label{eq:3}
  f(z)  = \frac{\sum_{j=0}^R\, n_j\, z^j}{1+\sum_{j=1}^{R'}\, d_j\, z^j}
\end{equation}
for known total degrees $R$ and $R'$.  Given a sequence of distinct
numbers $y_0,y_1,y_2,\ldots$, one can build a linear system of
equations for the coefficients $n_j$ and $d_j$ by evaluating the
function $f$ at $z=y_k$, namely
\begin{equation}
  \label{eq:ratrecsystem}
  \sum_{j=0}^R\, n_j\, y_k^j - \sum_{j=1}^{R'}\, d_j\, y_k^j\, f(y_k) = f(y_k).
\end{equation}
This strategy is even more convenient when a subset of the
coefficients is already known since it allows to significantly reduce
the number of needed evaluations of the function (this will also be
important later).

For the more general case where we do not have any information on the
degrees of the numerator and the denominator of the function, we use
Thiele's interpolation formula~\cite{abramowitz1964handbook},
\begin{align} \label{eq:thielerep}
  f(z) = {} & a_0 + \dfrac{z-y_0}{a_1 + \dfrac{z-y_1}{a_2 + \dfrac{z-y_3}{\cdots + \dfrac{z-y_{r-1}}{a_N}}}} \nn
       = {} & a_0 + (z-y_0)\left(a_1 + (z-y_1)\left( a_2 + (z-y_2) \left(\cdots+ \frac{z-y_{N-1}}{a_N}\right)^{-1} \right)^{-1} \right)^{-1},
\end{align}
where $y_0,y_1,\ldots$ is, once again, a sequence of distinct numbers.
Thiele's formula is the analogous for rational functions of Newton's
formula, and indeed it can be used in order to interpolate a
univariate rational function using the same strategy we illustrated
for the polynomial case.  Similarly as before, the result is converted
into a canonical form after the reconstruction.

The reconstruction of multivariate polynomials is performed by
recursively applying Newton's formula.  Indeed a multivariate
polynomial in $\z = \{z_1,\ldots,z_n\}$ can be seen as a univariate
polynomial in $z_1$ whose coefficients are multivariate polynomials in
the other variables $z_2,\ldots,z_n$,
\begin{equation} \label{eq:mnewtonrep}
  f(z_1,\ldots,z_n) ={} \sum_{r=0}^R a_r(z_2,\ldots,z_n) \prod_{i=0}^{r-1} (z_1-y_i).
\end{equation}
For any fixed numerical value of $z_2,\ldots,z_n$ one can apply the
univariate polynomial reconstruction algorithm in $z_1$ to evaluate
the coefficients $a_r$.  This means that the problem of reconstructing
an $n$-variate polynomial is reduced to the one of reconstructing an
$(n-1)$-variate polynomial.  Hence we apply this strategy recursively
until we reach the univariate case, which we already discussed.  The
result is then converted into the canonical form of
Eq.~\eqref{eq:poly}.

Before moving to the case of multivariate rational functions, it is
worth making a few observations on the choice of the sequence of
evaluation points $y_0,y_1,\ldots$ which appear in all the previous
algorithms.  We want to make a choice which does not interfere with
our capability of evaluating the function $f$ -- which may have
singularities both in its final expression and in intermediate stages
of its numerical evaluation -- and of inverting the relations for
obtaining Thiele's coefficients.  While making a choice which works
for any function is clearly impossible, in practice we can easily make
one which almost always works in realistic cases.  This is done by
choosing as $y_0$ a relatively large and random-like integer in
$\Z_p$, where common functions are extremely unlikely to have
singularities.  We then increase the integer by a relatively large
constant $\delta$ for the next points, i.e.\
$y_{i+1} = y_i + \delta \mod p$.  In the multivariate case, we use a
different starting point $y_0$ and a different constant $\delta$ for
each variable.  Heuristically we find that, with this strategy,
especially when using 64-bit primes, one can reasonably expect to find
no singular point even in millions of evaluations.

We finally discuss the more complex problem of reconstructing a
multivariate rational function $f=f(\z)$.  We first observe that the
reconstruction is much simpler when the constant term in the
denominator is non-vanishing since this unambiguously fixes the
normalization of the coefficients.  As suggested in
ref.~\cite{CUYT20111445}, we can force any function to have this
property by shifting its arguments by a constant vector
$\s=\{s_1,\ldots,s_n\}$ and reconstruct $f(\z+\s)$ instead.  In
practice, by default, we find it is convenient to always shift arguments
by a vector $\s$ such that any function coming from a realistic
problem is unlikely to be singular in $\z=\s$.  The criteria for the
choice of $\s$ are similar to the ones for choosing the sample points,
i.e.\ choosing relatively large and random-like numbers in $\Z_p$.
The result is shifted back to its original arguments after the full
reconstruction over a finite field $\Z_p$ is completed (note that this
detail differs from what is proposed in
ref.s~\cite{CUYT20111445,Peraro:2016wsq}).  Hence, in the following, we
assume that the function $f$ has a non-vanishing constant term in the
denominator, which by our choice of normalization is equal to one.

The key ingredient of the algorithm, which was also proposed in ref.~\cite{CUYT20111445}, is the introduction of an auxiliary variable
$t$ which is used to rescale all the arguments of $f$.  This defines
the function $h=h(t,\z)$, which takes the form
\begin{equation}
  \label{eq:hfunc}
  h(t,\z) \equiv f(t\, \z) = \frac{\sum_{r=0}^R\, p_r(\z)\, t^r}{1+\sum_{r=1}^{R'}\, q_r(\z)\, t^r}.
\end{equation}
In other words, $h(t,\z)$ is a univariate rational function in $t$,
whose coefficients $p_r$ and $q_r$ are multivariate homogeneous
polynomials of total degree $r$ in $\z$.  This allows to reconstruct
$f(\z)=h(1,z)$ by combining the algorithms discussed above for
univariate rational functions and multivariate polynomials.  In
practice, we start with a univariate reconstruction in $t$ for fixed
values of $\z$ using Thiele's formula, in order to get the total
degree of numerator and denominator.  This allows to check that the
denominator has indeed a non-vanishing constant term.  The knowledge
of the total degree also allows to use the system-solving strategy for
the next reconstructions in $t$.  We also perform a univariate
reconstruction of the unshifted function $f$ in each variable $z_j$
for fixed values of all the other variables.  The minimum degrees in
each variable are used to factor out polynomial prefactors in the
numerator and denominator of the function, which can significantly
reduce the number of required evaluations (note that it is essential
that this is done before shifting variables since after the shift any
realistic function is unlikely to have monomial prefactors).  The
maximum degrees are used to provide, together with the total degree
$r$ of each polynomial $p_r$ and $q_r$, the possibility of terminating
the polynomial reconstructions in the interested variables earlier.
They are also used in order to estimate a suitable set of sample
points for reconstructing the function before performing the
evaluations, as we will explain when discussing the parallelization
strategy in section~\ref{sec:parallelization}.  We then proceed with
using the system-solving strategy for univariate rational functions,
reconstructing $h(t,\z)$ as a function of $t$ for any fixed numerical
value of $\z$.  This provides an evaluation of the polynomials $p_r$
and $q_r$ at $\z$.  By repeating this for several values of $\z$, we
reconstruct these multivariate polynomials using Newton's formula
recursively.

A few observations are in order.  First, because the polynomials $p_r$
and $q_r$ are homogeneous, we can set $z_1=1$ and restore its
dependence at the end.  This makes up for having introduced the
auxiliary variable $t$.  Moreover, each reconstruction in $t$ provides
an evaluation of all the polynomial coefficients at the same time.
For this reason, for each $\z$ we cache the reconstructed coefficients
so that we can reuse the evaluations in several polynomial
reconstructions.  As for the reconstruction of the polynomials
themselves, we proceed from the ones with a lower degree to the ones of
higher degree (this detail is also different from what is presented in
ref.~\cite{Peraro:2016wsq}).  This way the polynomials with a lower
degree, which can be reconstructed with fewer evaluations, become
known earlier and can thus be removed from the system of equations in
Eq.~\eqref{eq:ratrecsystem} when reconstructing the ones with higher
degrees.  This makes the system of equations for higher-degree
polynomial coefficients smaller, and hence it further reduces the
number of needed evaluations.  As already mentioned, we also use the
information on the total degree $r$ of each polynomial, as well as the
maximum degree with respect to each variable, in order to terminate
the polynomial reconstructions earlier, when possible.

When combining all these ingredients, we find that the number of
evaluations we need for the reconstruction is comparable (if not
better) to the one we would need by writing a general ansatz based on
the degrees of the numerator and the denominator of the function (both
the total ones and the ones with respect to each variable).  However,
while the ansatz-based approach is impractical for complicated
multivariate functions since it requires to solve huge dense systems
of equations, the method presented here is instead able to efficiently
reconstruct very complex functions depending on several variables.  It
has indeed been applied to a large number of examples, some of which
have been mentioned in the introduction.

Finally, we point out that so far we only discussed single-valued
functions, but in the most common cases the output of an algorithm
will actually be a list of functions
\begin{equation}
  \label{eq:multivalf}
  \mathbf{f}(\z) = \{f_1(\z),f_2(\z),\ldots\}.
\end{equation}
In this case, the reconstruction proceeds as described above
considering one element of the output at the time.  However, for each
functional evaluation, the whole output, or a suitable subset of it, is
cached (more details on our caching strategy are discussed in
section~\ref{sec:proof-conc-impl}) so that the same evaluations can be
reused for the reconstruction of different functions $f_j(\z)$.

\subsection{Parallelization}
\label{sec:parallelization}

A well-known advantage of functional reconstruction techniques is the
possibility to extensively parallelize the algorithm.

The most important step which can be parallelized is the evaluation of
the function.  Since numerical evaluations are independent of each
other, they can be run in parallel over different threads, nodes, or
even on different machines.

Building an effective parallelization strategy is actually easier for
the multivariate case.  As discussed above, the multivariate
reconstruction begins with a univariate reconstruction in $t$ of the
function $h(t,\z)$ in Eq.~\eqref{eq:hfunc}, and univariate
reconstructions in each variable $z_j$ for fixed values of all the
other ones.  This amounts, for an $n$-variate problem, to $n+1$
univariate reconstructions, which being independent of each other can
be all run in parallel.  These univariate reconstructions are
significantly faster than a multivariate one and provide valuable
information for the multivariate reconstruction.  After this step we
use this information to determine a suitable set of evaluation points
for the reconstruction of each function in the output of the
algorithm, assuming the result is a generic function constrained by
the degrees found in the univariate fits.  While this might result in
building a set of evaluation points which is slightly larger than
needed, it allows to obtain a list of sample points which can be
independently evaluated before starting any multivariate
reconstruction.  Any performance penalty, due to this oversampling of
the function, is very small compared to what we gain from the
possibility of parallelizing the evaluations.  Therefore, this list of
points can be split according to the available computing resources and
evaluated in parallel over as many threads and cores as possible.

The main advantage of this parallelization strategy is the relative
ease of implementation since it requires minimal synchronization
(each thread just needs to evaluate all the points assigned to it and
wait for the others to finish), and the fact that it does not depend
on the specific numerical algorithm which is implemented.

After the evaluations have completed, they are collected and used for
the multivariate reconstruction algorithm described above, except that
the calls to the numerical ``black box'' procedure are now replaced by
a lookup of its values in the set of cached evaluations.

In building the list of evaluation points, one may initially assume
that a given number $n_p$ of primes will be needed for the
reconstruction over $\Q$ (typically, one will start with the choice
$n_p=1$), and that additional primes will only be used for a small
number of evaluations, for the purpose of checking the result.  If the
rational reconstruction described in section~\ref{sec:finite-fields}
fails these checks, points using additional primes will be added to
the list and another evaluation step will be performed for these.

We now turn to the univariate case.  Since building an effective and
clean parallelization strategy here is significantly harder than for
the multivariate case and in general, one does not need too many
evaluations for univariate reconstructions, by default we don't
perform any parallelization in the univariate case.  Indeed, the
strategy illustrated above would require us to perform a univariate
reconstruction over a finite field before performing any
parallelization, in order to obtain information on the degree of the
result.  For the univariate case, however, this task is of comparable
complexity than performing a complete reconstruction over $\Q$.
Despite this, if needed, we can still parallelize the evaluations in a
highly automated way as follows.  We start by making a guess on the
maximum degrees of numerator and denominator and build a set of
evaluation points based on this assumption.  Then we perform the
evaluations in parallel, as before.  After this, we proceed with the
reconstruction using the cached evaluations.  If during the
reconstruction we realize we need more evaluation points, we make a
more conservative guess of the total degrees and proceed again with
the evaluation (in parallel) of the additional points needed.  This
can be done automatically, by gradually increasing the maximum degree
by a given amount in each step.  We proceed this way until the
reconstruction is successful.  Obviously, making an accurate guess of
the total degrees may not be easy.  While making a conservative choice
of a high degree might result in too many evaluations, choosing a
total degree which is too low will cause the reconstruction to fail
and it will create additional overhead in launching the parallel tasks
for evaluating the additional points until the successful
reconstruction.  This method also requires some additional input from
the user of the reconstruction algorithm, which needs to provide these
guesses, since one cannot obviously make a choice which is good for
any problem.  For these reasons, we usually prefer to avoid
parallelization in univariate reconstruction, but it is still
important to know that a parallelization option is available for these
cases as well.

Another step which can be, to some extent, parallelized, is the
reconstruction itself.  As mentioned, in the most common cases, the
output of our algorithm is not a single function but a list of
functions.  Since the reconstructions of different functions from a
set of numerical evaluations are independent of each other, they can
also be run in parallel.  Even if this is generally not as important
as the parallelization of the functional evaluations, which are the
typical bottleneck, it can still yield a sizeable performance
improvement.

\section{Dataflow graphs}
\label{sec:dataflow-graphs}

In this section, we describe one of the main novelties introduced in
this paper, namely a method for building numerical algorithms over
finite fields using a special kind of computational graphs, known as
\emph{dataflow graphs}.

The algorithms described in the previous sections reduce the problem
of computing any (multivariate and multivalued) rational function to
the one of providing a numerical implementation of it, over finite
fields $\Z_p$.  The goal of the method described in this section is
providing an effective way of building this implementation,
characterized by good flexibility, performance, and ease of use.

\subsection{Graphs as numerical procedures}
\label{sec:graphs-as-numerical}

Dataflow graphs are directed acyclic graphs, which can be used to
represent a numerical calculation.  The graph is made of nodes and
arrows.  The \emph{arrows} represent data (i.e.\ numerical values in
our case) and the \emph{nodes} represent algorithms operating on the
(incoming) data received as input and producing (outgoing) data as
output.  In the following, we describe a simplified type of dataflow
graphs which we use in our implementation.

\begin{figure}[t]
  \centering
  \includegraphics[scale=0.4]{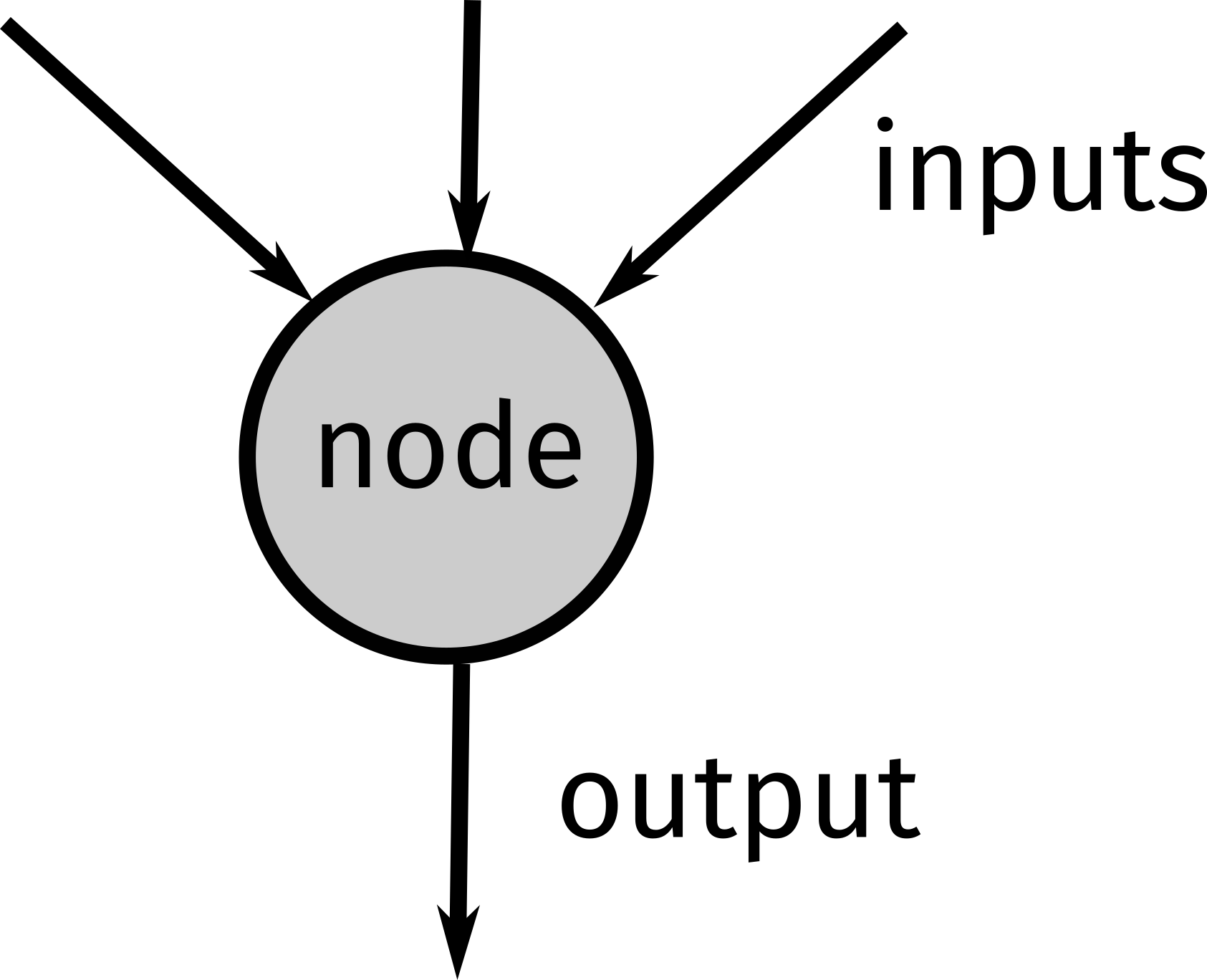}
  \caption{A node in a dataflow graph, where arrows represent lists of
    values and nodes represent numerical algorithms.  In our
    implementation, a node can take zero or more incoming arrows as
    input and has exactly one outgoing arrow as output.}
  \label{fig:node}
\end{figure}

In our case, an \emph{arrow} represents a \emph{list of values}.  A
\emph{node} represents a \emph{basic numerical algorithm}.  A node can
take zero or more incoming arrows (i.e.\ lists) as input and has
exactly one outgoing arrow\footnote{ In the graphical representations
  in this paper, if there are two or more outgoing arrows for a node,
  we understand that they all represent the same list.}
(i.e.\ one list) as output (see fig.~\ref{fig:node}).  For simplicity,
we also require that each list (represented by an arrow) has a well-defined length which cannot change depending on the evaluation point.
We also understand that nodes can also contain metadata with additional
information needed to define the algorithm to be executed.

Typically nodes encode common, basic algorithms (e.g.\ the evaluation
of rational functions, the solution of linear systems, etc\ldots)
which are implemented, once and for all, in a low-level language such
as \textsc{C++}.  We will give an overview of the most important ones
in section~\ref{sec:numer-algor-over}.  Complex algorithms are defined
by combining these nodes, used as building blocks, into a
computational graph representing a complete calculation, where the
output of a building block is used as input for others.  This way
complex algorithms are easily built without having to deal with the
low-level details of their numerical implementation.  The graph can
indeed be built from a high-level language, such as \textsc{Python} or
\textsc{Mathematica}.  Several explicit examples will be provided in
the next sections.

In each graph, there are two special nodes, namely the \emph{input
  node} and the \emph{output node}.  The input node does not represent
any algorithm, but only the list of input variables $\z$ of the graph.
The output node can be any node of the graph and represents, of
course, its output.  A dataflow graph thus defines a numerical
procedure which takes as input the variables $\z$ represented by the
input node and returns a list of values which is the output of the
output node.

Graphs are evaluated as follows.  First, every time we define a node,
we assign to it an integer value called \emph{depth}.  The depth of a
node is the maximum value of the depths of its inputs plus one.  The
depth of the input node is zero, by definition.  When an output node
is specified, we recursively select all the nodes which are needed as
inputs in order to evaluate it, and we sort this list by depth.  We
then evaluate all the nodes from lower to higher depths and store
their output values to be used as inputs for other nodes.  Once the
output node has been evaluated, its output is returned by the
evaluation procedure.

\subsection{Learning nodes}
\label{sec:learning-nodes}

As we already mentioned, each node has exactly one list of values as
output, and the length of this list is not allowed to change depending
e.g.\ on the evaluation point.  However, for some algorithms, we
cannot know the length of the output at the moment the numerical
procedure is defined.

Consider, as an example, a node which solves a linear system of
equations.  The length of the output of such a node depends on whether
the system is determined or undetermined and on its rank.  This
information is usually not known a priori but it must be
\emph{learned} after the system is defined.  In this case, it can
easily be learned by solving the system numerically a few times.

For this reason, we allow nodes to have a \emph{learning phase}.  The
latter is algorithm-dependent and typically consists of a few
numerical evaluations used by a node in order to properly define its
output.  Hence, the output of these nodes can be used as input by
other nodes only after the learning phase is completed (since, before
that, their output cannot be defined at all).

More algorithms which require a learning phase will be discussed
later.

\subsection{Subgraphs}
\label{sec:subgraphs}

An important feature which makes this framework more powerful and
usable in realistic problems is the possibility of defining nodes in
which one can embed other graphs.

Consider a graph $G_1$ with a node $N$ which embeds a graph $G_2$.  We
say that $G_2$ is a \emph{subgraph}.  Typically, the node $N$ will
need to evaluate the subgraph $G_2$ a number of times in order to
produce its output.

The simplest case of a subgraph is when the node $N$ takes one list as
input, passes the same input to $G_2$ in order to evaluate it, and
then returns the output of $G_2$.  This case, which we call
\emph{simple subgraph}, is equivalent to having the nodes of $G_2$
attached to the input node of $N$ inside the graph $G_1$ directly, but
it can still be useful in order to organize more cleanly some
complicated graphs.

Another interesting example, which we call \emph{memoized subgraph},
can be beneficial when parts of the calculation are independent of
some of the variables.  This type of subgraph effectively behaves the
same way as the simple subgraph described above, except that it
remembers the input and the output of the last evaluation.  If the
subgraph needs to be evaluated several times in a row with the same
input, the memoized subgraph simply returns the output it has stored.
This is particularly useful when combined with the Laurent expansion,
the subgraph fit, or the subgraph multi-fit algorithms.  We will give
a description of these later in this paper, but for now, it suffices to
know that they require to evaluate a dataflow graph several times for
fixed values of a subset of the variables.  In such cases, one may not
wish to evaluate every time the parts of a graph which only depend on
the variables which remain fixed for several evaluations.  One can
thus optimize away these evaluations by embedding the appropriate
parts of the graph in a memoized subgraph.

One more useful type of subgraph is a \emph{subgraph map}.  This takes
an arbitrary number of lists of length $n$ as input, where $n$ is the
number of input parameters for $G_2$.  The graph $G_2$ is then
evaluated for each of the input lists, and the outputs are chained
together and returned.  This is useful when the same algorithm needs
to be evaluated for several inputs.

There are however other interesting cases, where the node $N$ requires
to evaluate $G_2$ several times and perform non-trivial operations on
its output.  Some useful examples are given at the end of
section~\ref{sec:linear-fit} and in
section~\ref{sec:laurent-expansion}.

\section{Numerical algorithms over finite fields}
\label{sec:numer-algor-over}

In this section, we discuss several basic, numerical algorithms which
can be used as nodes in a graph.  These are best implemented in a
low-level language such as \textsc{C++} for efficiency reasons.  In
later sections, we will then show how to combine these basic building
blocks into more complex algorithms which are relevant for
state-of-the-art problems in high-energy physics.

\subsection{Evaluation of rational functions}
\label{sec:eval-polyn-rati}

Most of the algorithms we are interested in have some kind of analytic
input, which can be cast in the form of one or more lists of
polynomials or rational functions.  The numerical evaluation of
rational functions is, therefore, one of the most ubiquitous and
important building blocks in our graphs.
\begin{figure}[t]
  \centering
  \includegraphics[scale=0.35]{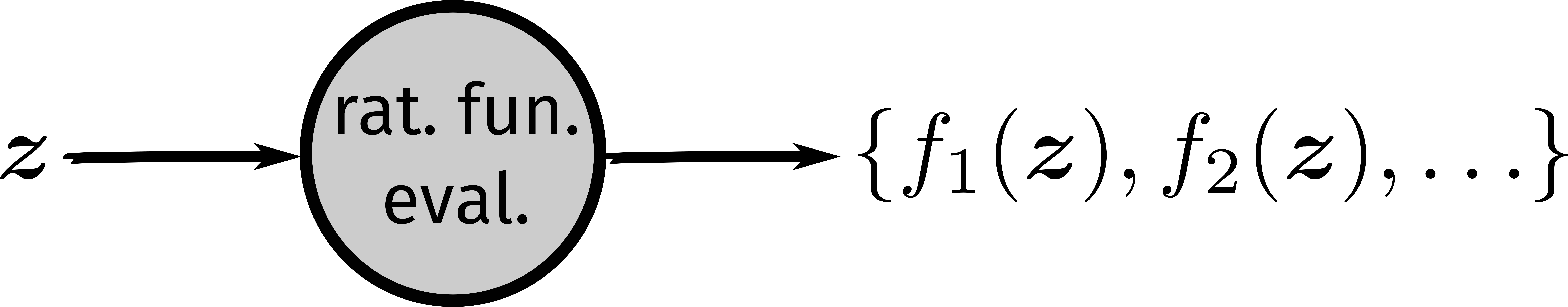}
  \caption{A node representing the evaluation of a list of rational
    functions.  It takes one list $\z$ as input, interpreted as the
    list of arguments of the functions, and returns the values of the
    functions evaluated at $\z$.}
  \label{fig:funceval}
\end{figure}
These nodes take as input one list of values $\z$ and return a list of
rational functions $\{f_j(\z)\}$ evaluated at that value, as
schematically illustrated in fig.~\ref{fig:funceval}.

Polynomials are efficiently evaluated using the well known Horner
scheme.  Given a univariate polynomial
\begin{equation}
  \label{eq:unipoly}
  p(z) = \sum_{j=0}^R\, c_j\, z^j,
\end{equation}
Horner's method is based on expressing it as
\begin{equation}
  \label{eq:hornerpoly}
  p(z) = c_0 + z\, \left(c_1 + z\, \left( c_2 + \cdots z\, \left(c_{R-1} + z\, c_R \right) \right)  \right).
\end{equation}
This formula only has $R$ multiplications and $R$ additions for a
polynomial of degree $R$, and it can be easily obtained from the
canonical representation in Eq.~\ref{eq:unipoly}.  Therefore, it is a
great compromise between ease of implementation and efficiency.

For multivariate polynomials, Horner's scheme is applied recursively
in the variables.  In practice, we use an equivalent but non-recursive
implementation and we store all the polynomial data (i.e.\ the integer
coefficients $c_j$ and the metainformation about the total degrees of
each sub-polynomial) in a contiguous array of integers.

Rational functions are obviously computed as ratios of two
polynomials.  If the denominator vanishes for a specific input, the
evaluation fails and yields a singular point.

\subsection{Dense and sparse linear solvers}
\label{sec:dense-sparse-linear}

A wide variety of algorithms involves solving one or more linear
systems at some stage of the calculation.  Moreover, the solution of
these systems is often the main bottleneck of the procedure, hence
having an efficient numerical linear solver is generally very important.

In general, consider a $n\times m$ \emph{linear system} with
parametric rational entries in the parameters $\z$,
\begin{equation}
  \sum_{j=1}^m A_{ij}\, x_j = b_i, \quad (i=1,\ldots, n),
\end{equation}
with
\begin{equation}
  \label{eq:6}
  A_{ij} = A_{ij}(\z), \quad b_i=b_i(\z).
\end{equation}
This is defined by the matrix $A=A(\z)$, the vector $b=b(\z)$, and the
set of $m$ variables or unknowns $\{x_j\}$.  We assume there is a
total ordering between the unknowns,
$x_1\succ x_2 \succ \cdots \succ x_m$.  Borrowing from a language
commonly used in the context of IBP identities, we say that $x_1$ has
higher \emph{weight} than $x_2$ and so on.  This simply means that,
while solving the system, we always prefer to write unknowns with
higher weight in terms of unknowns with a lower weight.

\begin{figure}[t]
  \centering
  \includegraphics[scale=0.35]{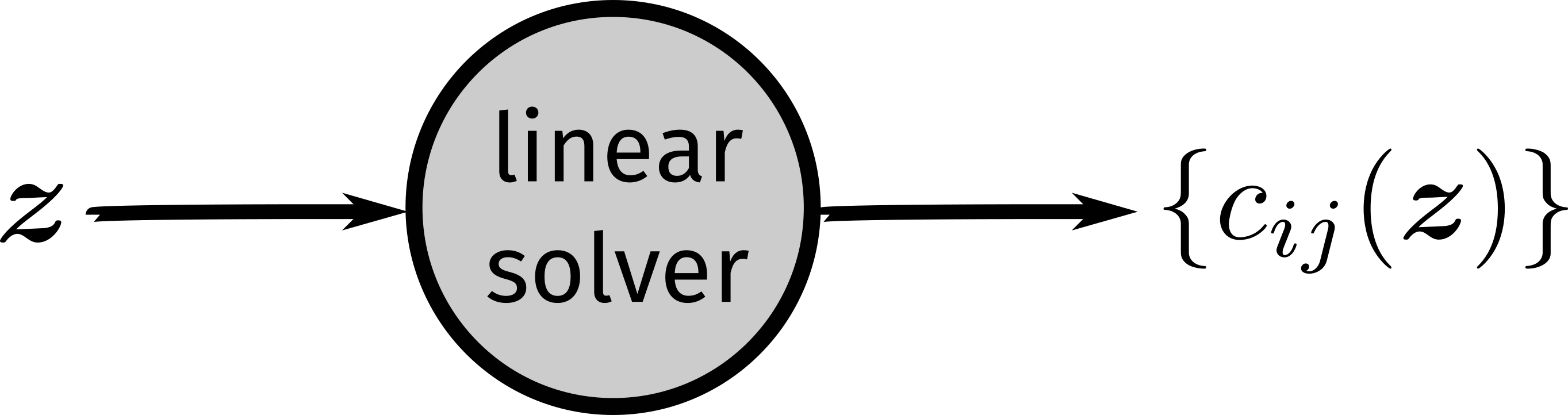}
  \caption{A node representing a linear solver.  It takes as input a
    list of parameters $\z$ and returns the coefficients of the
    solution defined in Eq.~\eqref{eq:linsolsol}
    and~\eqref{eq:linsolcmat}.}
  \label{fig:linsol}
\end{figure}

For each numerical value of $\z$ and prime $p$, the entries
$A_{ij}(\z)$ and $b_i(\z)$ are evaluated and the numerical system is
thus solved over finite fields.  If the system is determined, for each
numerical value of
$\z$ the solver returns a list of values for the
unknowns $x_j$.  In the more general case where there are fewer
independent equations than unknowns, one can only rewrite a subset of
the unknowns as linear combinations of others.  This means that we
identify a subset of \emph{independent} unknowns and the
complementary subset of \emph{dependent} unknowns which are written as
linear combinations of the independent ones,
\begin{equation}
  \label{eq:linsolsol}
  x_i = \sum_{j\in\text{indep.}} c_{ij}\, x_j + c_{i0}\qquad (i\in \text{dep.}).
\end{equation}
Notice that the list of dependent and independent unknowns also
depends on the chosen ordering (or weight) of the unknowns.  The
output of a linear solver is a list with the coefficients $c_{ij}$
appearing in this solution.  More specifically they are the rows of
the matrix
\begin{equation}
  \label{eq:linsolcmat}
\left[
\begin{array}{c|c}
\{ \vphantom{A p} c_{ij}  \}_{j\in\text{indep.}} & c_{i0} \\
\end{array}
\right]_{i\in \text{dep.}}
\end{equation}
stored in row-major order.  If only the homogeneous part of the
solution is needed, the elements $c_{i0}$ are removed from the output.
A node representing a linear solver is schematically depicted in
fig.~\ref{fig:linsol}.

It often happens that only a subset of the unknowns of a system is
actually needed.  We therefore also have the possibility of optionally
specifying a list of \emph{needed unknowns}.  When this is provided,
only the part of the solution which involves needed unknowns on the
left-hand side is returned.  This also allows to perform some further
optimizations during the solution of the system, as we will show
later.

As mentioned in section~\ref{sec:learning-nodes}, a linear solver is
an algorithm which needs a \emph{learning step}.  During this step,
with a few numerical solutions of the system, the list of dependent
and independent unknowns is learned.  This step is also used in order
to identify redundant equations, i.e.\ equations which are reduced to
$0=0$ after the solution, which are thus removed from the system,
improving the performance of later evaluations.  Moreover, the list of
dependent and independent unknowns is checked during every evaluation
against the one obtained in the learning step, since accidental zeroes
may change the nature of the solution of the system.  If the two do
not agree, the evaluation fails and the input is treated as a singular
point.

It is useful to distinguish between \emph{dense} and \emph{sparse}
systems of equations.  Even if they represent the same mathematical
problem, from a computational point of view they are extremely
different.

\emph{Dense systems} of equations are systems where most of the
entries in the matrix $A$ defined above are non-zero.  For these
systems, we store the $n$ rows of the matrix
\begin{equation}
  \label{eq:linsolABmat}
\left[
\begin{array}{c|c}
\vphantom{A p} A & b \\
\end{array}
\right]
\end{equation}
as contiguous arrays of $m+1$ integers.  We also add an $(m+2)$-th
entry to these arrays which assigns a different numerical ID to each
equation, for bookkeeping purposes.  The solution is a straightforward
and rather standard implementation of Gauss elimination.  This
distinguishes two phases.  The first, also known as forward
elimination, puts the system in \emph{row echelon form}.  The second,
also known as back substitution, effectively solves the systems by
putting it in \emph{reduced row echelon form}.  The algorithm we use
for dense systems works as follows.
\begin{description}
\item[Forward elimination] We set a counter $r=0$, and loop over the
  unknowns $x_k$ for $k=1,\ldots,m$, i.e.\ from higher to lower
  weight.  At iteration $k$, we find the first equation $E_j$ with
  $j\geq r$ where the unknown $x_k$ is still present.  If there is no
  such equation, we move on to the next iteration.  Otherwise, we move
  equation $E_j$ in position $r$, and we ``solve'' it with respect to
  $x_k$, i.e.\ we normalize it such that $x_k$ has coefficient equal
  to 1.  We thus substitute the equation in all the remaining
  equations below position $r$.  We then increase $r$ by one and
  proceed with the next iteration.
\item[Back substitution] We loop over the equations $E_r$, for
  $r=n,n-1,\ldots,1$, i.e.\ from the one in the last position to the
  one in the first position.  At iteration $r$, we find the highest
  weight unknown $x_j$ appearing in equation $E_r$ (note that this is
  guaranteed to have coefficient equal to one, after the forward
  elimination).  If equation $E_r$ does not depend on any unknown, we
  proceed to the next iteration.  Otherwise, we substitute equation
  $E_r$, which contains the solution for $x_j$, in all the equations
  $E_k$ with $k<r$.  We then proceed with the next iteration in $r$.
\end{description}
During the learning phase, the system is solved in order to learn
about the independent variables and the independent equations.  The
remaining equations, once the system has been reduced, will become
trivial (i.e.\ $0=0$) and will, therefore, be removed.  We also identify
unknowns which are zero after the solution.  These are then removed
from the system and this allows to find, through another numerical
evaluation, a smaller set of independent equations needed to solve for
the non-zero unknowns.  We recall that solving a dense $n\times n$
system has $\O(n^3)$ complexity, hence it scales rather badly with the
number of equations and unknowns, and it greatly benefits from the
possibility of removing as many equations as possible.

We now discuss the reduction of \emph{sparse} systems of equations,
i.e.\ systems where most of the entries of the matrix $A$ which
defines it are zero.  In other words, in such a system, most of the
equations only depend on a relatively small subset of variables.  We
represent sparse systems using a sparse representation of the rows of
the matrix~\eqref{eq:linsolABmat}.  More specifically, for each row, we
store a list of non-vanishing entries, with the number of their
columns and their numerical value.  These are always kept sorted by
column index from the lowest to the highest, or equivalently by the
weight of the corresponding unknown from the highest to the lowest.
We also store additional information, namely the number of
non-vanishing terms in the row, and the index of the equation
corresponding to that row.  When solving such systems, it is crucial
to keep the equations as simple as possible at every stage of the
solution.  This way the complexity of the algorithm can have a much
better scaling behaviour than the one which characterizes dense
systems (the exact scaling strongly depends on the system itself, and
it can be as good as $\O(n)$ in the best scenarios, and as bad as
$\O(n^3)$ in the worst ones).  For these reasons, we implement a
significantly different version of Gauss elimination for sparse
systems, which shares many similarities with the one illustrated
in~\cite{Maierhoefer:2017hyi}.  We first sort the equations by
complexity, from lower to higher.  The complexity of an equation is
defined the same way as in ref.~\cite{Maierhoefer:2017hyi}, and is
determined by the following criteria,
sorted by their importance,
\begin{itemize}
\item the highest weight unknown in the equation (higher weight means
  higher complexity)
\item the number of unknowns appearing in the equation (a higher
  number means higher complexity)
\item the weight of the other unknowns in the equation, from the ones
  with higher weight to the ones with lower weight (i.e.\ if two
  equations have the same number of unknowns, the most complex one is
  the one with the highest weight unknown among those that are not
  shared by both equations).
\end{itemize}
If all the three points above result in a tie between two equations,
it means that they depend on exactly the same list of unknowns, and we
say that they are equally complex, hence their relative order does not
matter.  Obviously, other more refined definitions of this complexity
are possible, but we find that this one works extremely well for
systems over finite fields, despite its simplicity.  Once the
equations are sorted, the algorithm for sparse systems works as
follows for the forward and back substitution.
\begin{description}
\item[Forward elimination] We create an array $S$ whose length is
  equal to the number of unknowns.  This will contain, at position
  $j$, the index $S(j)$ of the equation containing the solution for
  the unknown $x_j$, or a flag indicating that there is no such
  equation.  We loop over the equations $E_i$ for $i=1,\ldots,n$, from
  lower to higher complexity.  If any equation is trivial (i.e.\
  $0=0$), we immediately move to the next one.  We find the unknowns
  appearing in $E_i$ for which a solution is already found, via
  lookups in the array $S$.  Among these, we select the unknown $x_k$
  such that $E_{S(k)}$ has the lowest complexity.  The equation
  $E_{S(k)}$ is then substituted into $E_i$.  This is repeated until
  all the unknowns in $E_i$ have no solution registered in $S$.  We
  then take the highest weight unknown $x_h$ and ``solve'' the
  equation with respect to it.  Once again, this means that we
  normalize the equation such that the coefficient of $x_h$ is one.
  We then register this solution in $S$ by setting $S(h) = i$, and
  proceed with the next iteration in $i$.
\item[Back substitution] We remove from the system any equation which
  has become trivial ($0=0$), but otherwise, we keep them in the same
  order.  We also update the array $S$ to take this change into
  account.  Let thus $n_I$ be the number of independent equations
  which survived after the forward elimination.  We loop again over
  the remaining equations $E_i$ for $i=1,\ldots,n_I-1$, from lower to
  higher complexity, excluding the last one.  If a list of needed
  unknowns was specified, and the highest weight unknown in equation
  $E_i$ is not in it, the equation is skipped.  We then find the
  unknowns in $E_i$, excluding the highest weight one, for which a
  solution is registered in $S$.  Among these, we pick the unknown
  $x_k$ such that the equation $E_{S(k)}$ has the lowest complexity.
  We then substitute equation $E_{S(k)}$ in $E_i$.  This is repeated
  until none of the unknowns in $E_i$, except the one with the highest
  weight, has a registered solution in $S$.
\end{description}
Similarly as before, during the learning step we identify the
independent equations, removing all the other ones, and the
independent unknowns.  For each equation $E_i$, we also keep track of
all the other equations which have been substituted into $E_i$ either
during the forward elimination or the back substitution.  This
information can optionally be used in order to further reduce the
number of needed equations.  Indeed, while after the learning stage
the system is guaranteed to contain only independent equations, there
might be a smaller subset of them which is still sufficient in order
to find a solution for all the \emph{needed unknowns}, which sometimes
are a significantly smaller subset of the ones appearing in the
system.  This simplification is obtained, when requested, by means of
the \emph{mark and sweep} algorithm.\footnote{The mark-and-sweep
  method is a well known algorithm primarily used for automatic memory
  management (garbage collection) in order to reclaim the unused
  memory of a computer program.  Here we use it instead to identify
  equations which are no longer useful (rather than allocated memory
  which is no longer being used), but it is based on the same
  mechanism.}  After the learning stage, for each equation $E_j$ we
have a list of dependencies $L_j$.  If $E_k\in L_j$, then $E_j$
depends on $E_k$, because $E_k$ was substituted into $E_j$ at some
point during the Gauss elimination.  We identify a set $R$ containing
the so-called \emph{roots}, which in our case are the equations
containing solutions for the needed unknowns.  We then ``mark'' all
the equations in $R$.  ``Marking'' is a recursive operation achieved,
for any equation $E_j$, by setting a flag which says that $E_j$ is
needed, and then recursively marking all the equations in $L_j$ whose
flag hasn't already been set.  Finally, we ``sweep'', i.e.\ discard
all the equations which have not been marked.  Notice that the mark
and sweep algorithm loses some information about the system, and
therefore it is only performed upon request.  It is however extremely
useful, e.g.\ when solving IBP identities, since it often reduces the
size of the system by a factor even greater than the simplification
achieved in the learning stage.

We also implement a dense solver algorithm called \emph{node dense
  solver}, which takes the elements of the matrix in
Eq.~\eqref{eq:linsolABmat} from its input node, in row-major order,
rather than from analytic formulas.  In the future, we may implement a
\emph{node sparse solver} as well, which only takes the non-vanishing
elements of that matrix from its input node, and uses a sparse solver
for the solution.

It goes without saying that these linear solvers can also be used in
order to invert matrices, using the Gauss-Jordan method.  Indeed, the
inverse of a $n\times n$ matrix $A_{ij}$ is the output of a linear
solver node which solves the system
\begin{equation}
  \label{eq:inverse}
  \sum_{j=1}^n A_{ij} x_j - t_i = 0, \qquad i=1,\ldots,n,
\end{equation}
with respect to the following unknowns, sorted from higher to lower
weights,
\begin{equation*}
  \{x_1,\ldots,x_n,t_1,\ldots,t_n\}.
\end{equation*}
In particular, when only the homogeneous part of the solution is
returned, the output of such a node will be a list with the matrix
elements $A^{-1}_{ij}$ in row-major order.  Both the dense and the
sparse solver can be used for this purpose, depending on the sparsity
of the matrix $A_{ij}$.  Also, notice that the matrix $A_{ij}$ is
invertible if and only if $\{x_j\}$ is the list of dependent unknowns
and $\{t_j\}$ is the list of independent unknowns.  This can be
checked after the learning phase has completed.

\subsection{Linear fit}
\label{sec:linear-fit}

Linear fits are another important algorithm which is often part of
calculations in high energy physics.  For instance, it is the main
building block of integrand reduction methods (see
section~\ref{sec:integrand-reduction}).  They are also used, for
instance, in order to match a result into an ansatz and to find
linear relations among functions.

In general, in a linear fit, we have two types of variables, which
in this section we call $\z=\{z_j\}$ and $\mathbf{\tau}=\{\tau_j\}$.
In particular, the $\z$ variables are simply regarded as free
parameters.  A linear fit is thus defined by an equation of the form
\begin{equation}
  \label{eq:linfit}
  \sum_{j=1}^m x_j(\z)\, f_j(\mathbf{\tau},\z) + f_0(\mathbf{\tau},\z) = g(\mathbf{\tau},\z)
\end{equation}
where $f_j$ and $g$ are \emph{known} (or otherwise \emph{computable})
rational functions and the coefficients $x_j$ are \emph{unknown}.
While $f_j$ and $g$ depend on both sets of variables, the unknown
coefficients $x_j$ can depend on the free parameters $\z$ only.  For
each numerical value of $\z$, Eq.~\eqref{eq:linfit} is sampled for
several numerical values of the variables $\mathbf{\tau}$.  This will
generate a linear system of equations for the unknowns $x_j$.  Linear
fits are thus just a special type of dense linear systems.  Hence, we
refer to the previous section for information about the implementation
of the reduction and the output of this algorithm.  In particular,
each equation is associated with a particular numerical sample point for
the variables $\mathbf{\tau}=\{\tau_j\}$.  In total, we use
$m+n_{\textrm{checks}}$ sample points, where $m$ is the number of
unknowns and $n_{\textrm{checks}}$ is the number of additional
equations added as a further consistency check (we typically use
$n_{\textrm{checks}}=2$).  Notice that, just like in any other linear
system, redundant equations (including the additional
$n_{\textrm{checks}}$ ones) are eliminated after the learning phase.

In order to use this algorithm more effectively for the solution of
realistic problems, and in particular integrand reduction, we made it
more flexible by adding some additional features.  The first one is
the possibility of introducing a set of auxiliary functions
$\mathbf{a}=\mathbf{a}(\mathbf{\tau},\z)$ and defining several (known)
functions $g_j$ on the right-hand side, in order to rewrite
Eq.~\eqref{eq:linfit} as
\begin{equation}
  \label{eq:linfitaux}
  \sum_{j=0}^m x_j(\z)\, f_j(\z, \mathbf{a}(\mathbf{\tau},\z)) + f_0(\z, \mathbf{a}(\mathbf{\tau},\z)) = \sum_j w_j\, g_j(\z, \mathbf{a}(\mathbf{\tau},\z)).
\end{equation}
This is useful when the functions $f_j$ and $g_j$ are simpler if
expressed in terms of these auxiliary variables $\mathbf{a}$, which do
not need to be independent, and when the sum on the right-hand side is
not collected under a common denominator.  The value of the
\emph{weights} $w_j$ in the previous equation depends on the inputs of
the node defining the algorithm.  The first input list is always the
list of variables $\z$, similarly to the case of a linear system.  If
no other input is specified, then we simply define $w_j=1$ for all
$j$.  If other lists of inputs are specified, besides $\z$, they are
joined and interpreted as the weights $w_j$ appearing in
Eq.~\eqref{eq:linfitaux}.  This allows to define these weights
numerically from the output of other nodes.  As we will see in
section~\ref{sec:integrand-reduction}, this allows, among other
things, to easily implement multi-loop integrand reduction over finite
fields without the need of writing any low-level code.

We provide two more usages of linear fits as nodes embedding
\emph{subgraphs}\ (introduced in section~\ref{sec:subgraphs}).  The
first one is used to find linear relations among the entries of the
output of the subgraph $G$ which has input variables
$\{\mathbf{\tau},\z\}$.  Let
\begin{equation}
  \label{eq:Goutputfj}
  \{f_1(\mathbf{\tau},\z),\ldots, f_m(\mathbf{\tau},\z)\}
\end{equation}
be the output of $G$.  The \emph{subgraph fit} algorithm solves the
linear fit problem
\begin{equation}
  \label{eq:11}
  \sum_{j=1}^{m-1} x_j(\z)\, f_j(\mathbf{\tau},\z) = f_m(\mathbf{\tau},\z).
\end{equation}
In particular, if $\z$ is chosen to be the empty list, and $f_m=0$, it
will find vanishing linear combinations of the output of $G$ with
numerical coefficients.  An interesting application of this is the
attempt of simplifying the output of a graph.  One can indeed estimate
the complexity of each entry in the output at the price of relatively
quick univariate reconstructions.  A simple way of estimating the
complexity is based on the total degrees of numerators and
denominators, which can be found with one univariate reconstruction
over one finite field, as we already explained.  A more refined method
would be counting the number of evaluation points needed for the
reconstruction of each entry over a finite field, which can be found
after the total and partial degrees have been computed and it is an
upper bound on the number of non-vanishing terms in the functions.
One can, of course, use any other definition or estimate for the
complexity of the output functions based on other elements specific to
the considered problem.  Regardless of how we choose to define it, we
then sort the entries by their complexity, from lower to higher, and
we make sure that $f_m=0$, e.g.\ by appending to the graph a Take node
(this will be described in section~\ref{sec:basic-oper-lists}).  After
solving the linear fit above for the unknowns $x_j$ we are then able
to write more complex entries of the output as linear combinations of
simpler entries.  When this is possible, only the independent entries
need to be reconstructed.

The second subgraph application of linear fits, which we call
\emph{subgraph multi-fit}, is a generalization of the previous one.
If Eq.~\eqref{eq:Goutputfj} represents, again, the output of a graph
$G$, the subgraph multi-fit node, which has input variables $\z$, is
defined by providing a list of lists of the form
\begin{equation}
  \label{eq:15}
  \{\{\sigma_{1j}\}_{j=1}^{l_1},\{\sigma_{2j}\}_{j=1}^{l_2},\ldots\}
\end{equation}
where the sublists can be of any length and $\sigma_{ij}$ are integer
indexes in the interval $[1,m]$.  For each sublist
$\{\sigma_{ij}\}_j$, the subgraph multi-fit node solves the linear fit
\begin{equation}
  \label{eq:16}
  \sum_{j=1}^{l_i-1}\, x_{ij}(\z)\, f_{\sigma_{ij}}(\mathbf{\tau},\z) = f_{\sigma_{i\, l_i}}(\mathbf{\tau},\z),
\end{equation}
with respect to the unknowns $x_{ij}$.  Since this amounts to
performing a number of linear fits, this node obviously has a learning
phase, where independent unknowns, independent equations, and zero
unknowns are detected for each one of them.  Notice that all the fits
can share the same evaluations of graph $G$, for several values of
$\tau$ and fixed values of $\z$.  An application of this algorithm is
the case when the functional dependence of a result on the subset of
variables $\mathbf{\tau}$ (which may also be the full set of
variables, if $\z$ is the empty list) can be guessed a priori by
building a basis of rational functions.  In this case, one may create
a graph $G$ which contains both the result to be reconstructed and the
elements of the function basis, and a second graph with a subgraph-fit
node using $G$ as a subgraph.  This allows to reconstruct the result
via a simpler functional reconstruction over the $\z$ variables only,
or via a numerical reconstruction if $\z$ is the empty list.  An
example of this is given at the end of
section~\ref{sec:diff-equat-epsil}.

\subsection{Basic operations on lists and matrices}
\label{sec:basic-oper-lists}

The algorithms listed in this subsection have a simple implementation
and they can be thought as utilities for combining in a flexible way
outputs of other numerical algorithms in the same graph.  While they
typically execute very quickly compared to others, they greatly
enhance the possibilities of defining complex algorithms by means of
the graph-based approach described in this paper.  They are:
\begin{description}
\item [Take] Takes any number of lists as input and returns a
  specified list of elements $\{t_1,t_2,\ldots\}$ from them, where
  $t_j$ can be any element of any of the input lists.  The same
  element may also appear more than once in the output list.  This is
  a very flexible algorithm for rearranging the output of
  (combinations of) other nodes.  Indeed many of the list-manipulation
  algorithms below can also be implemented as special cases of this.
\item [Chain] Takes any number of lists as input, chains them and
  return them as a single list.
\item [Slice] Takes a single list as input and returns a slice (i.e.\
  a contiguous subset of it) as output.
\item [Matrix Multiplication] Given three positive integers $N_1$,
  $N_2$ and $N_3$, this node takes two lists as input, interprets them
  as the entries of a $N_1\times N_2$ matrix and a $N_2\times N_3$
  matrix (in row-major order) respectively, multiplies them and return
  the entries of the resulting $N_1\times N_3$ matrix (still in
  row-major order).  This node is depicted in fig.~\ref{fig:matmul}.
\begin{figure}[t]
  \centering
  \includegraphics[scale=0.4]{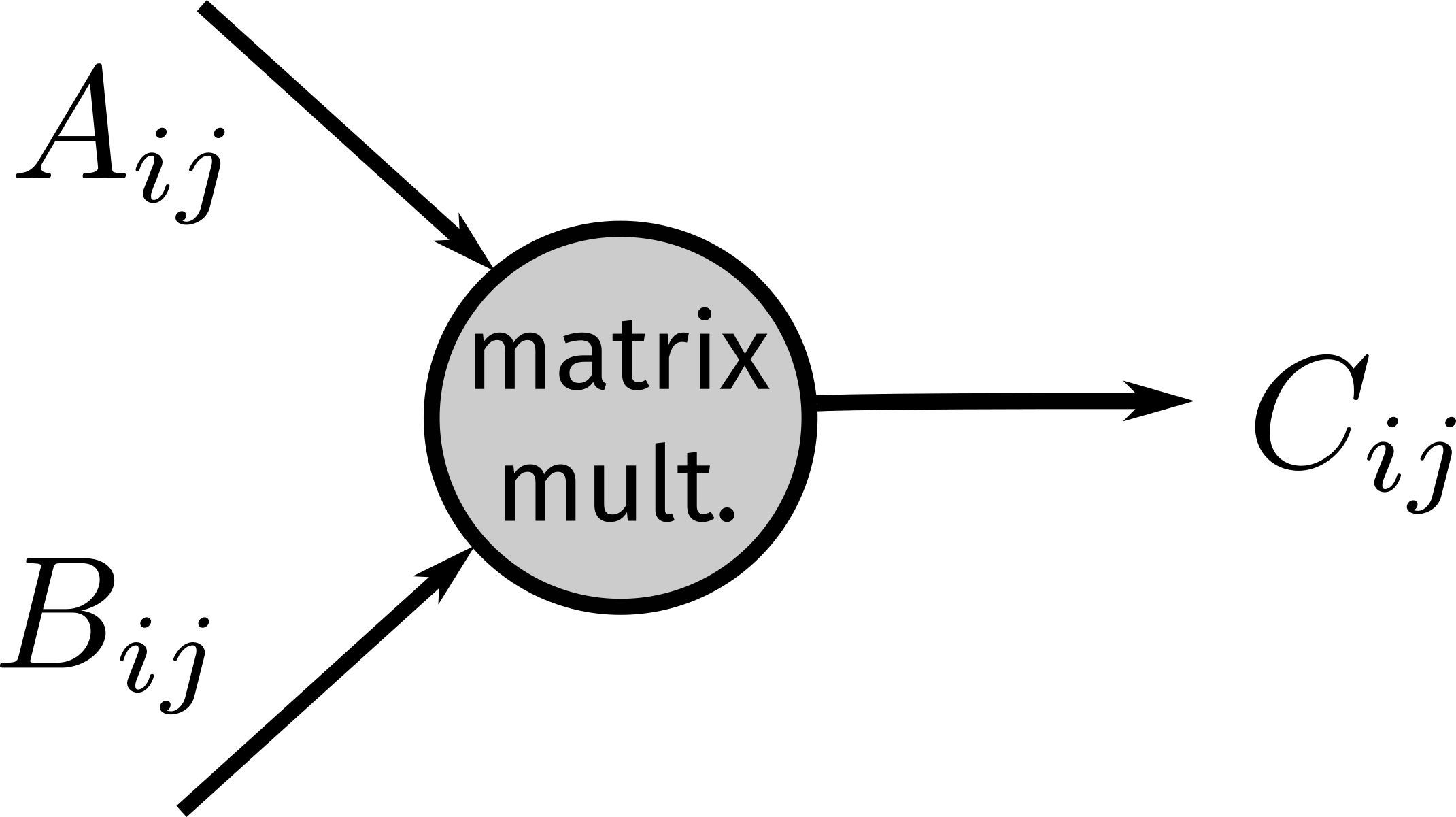}
  \caption{A node representing the matrix multiplication
    $C_{ij}=\sum_k A_{ik}\, B_{kj}$.  The arrows represent lists with
    the matrix elements $A_{ij}$, $B_{ij}$ and $C_{ij}$ in row-major
    order.  Their number of rows and columns is defined when the
    node is created.}
  \label{fig:matmul}
\end{figure}
  Notice that, because different nodes of this type
  can interpret the same inputs as matrices of different sizes (as long
  as the total number of entries is consistent), this algorithm can
  also be used to contract indexes of appropriately stored tensors,
  multiplying lists and scalars, and other similar operations.  As an
  example, consider a node whose output are the entries of a rank 3
  tensor $T_{A B C}$ with dimensions $N_A$, $N_B$, $N_C$, and another
  one which represents a matrix $M_{CD}$ with dimensions $N_C$ and
  $N_D$.  We can then perform a tensor-matrix multiplication using
  this node with $N_1= N_A\times N_B$, $N_2=N_C$ and $N_3=N_D$.
  Similarly, we can multiply the tensor $T_{A B C}$ by a scalar, the
  latter represented by a list of length one, by setting
  $N_1= N_A\times N_B\times N_C$, $N_2=1$, $N_3=1$.
\item [Sparse Matrix Multiplication] Similar to the Matrix
  Multiplication above, but more suited for cases where the matrices
  in the input are large and sparse, so that one wants to store only
  their non-vanishing entries in the output of a node.  This algorithm
  is defined by the three dimensions $N_1$, $N_2$ and $N_3$ as above,
  as well as by a list of potentially non-vanishing columns for each
  row of the two input matrices.  The two inputs are then interpreted
  as lists containing only these potentially non-vanishing elements.
  The output of this node lists, as before, the elements of the
  resulting $N_1\times N_3$ matrix, stored in a dense representation
  in row-major order.
\item [Addition] Takes any number of input lists of length $L$, and
  adds them element-wise.
\item [Multiplication] Takes any number of input lists of length $L$, and
  multiplies them element-wise.
\item [Take And Add] Similar to the Take algorithm above, except that
  it takes several lists from its inputs
  $\{\{t_{1j}\}_j,\{t_{2j}\}_j,\ldots\}$, where each of them might
  have a different length.  It then returns the sum of each of these
  sub-lists $\{\sum_j t_{1j},\sum_jt_{2j},\ldots\}$.
\item [Non-Zeroes] This node takes one list as input and returns only
  the elements which are not identically zero.  The node requires a
  learning step where the non-zero elements are identified via a few
  numerical evaluations (two by default).  Because some algorithms
  have a rather sparse output (i.e.\ with many zeroes), it is very
  often useful to append this node at the end of a graph and use it as
  output node.  This can remarkably improve memory usage during the
  reconstruction step.  Given its benefits and its minimal impact on
  performance, we also recommend using such an algorithm as the output
  node when the sparsity of the output is not known a priori.
\end{description}

\subsection{Laurent expansion}
\label{sec:laurent-expansion}

In physical problems, one is often interested in the leading
coefficients of the Laurent expansion of a result with respect to one
of its variables, which in this section we call $\epsilon$.  The most
notable examples in high-energy physics are scattering amplitudes in
dimensional regularization, which are expanded for small values of the
dimensional regulator.  Other applications can be the expansion of a
result around special kinematic limits.  The coefficients of this
expansion are often expected to be significantly simpler than the full
result.  Hence, it is beneficial to be able to compute the Laurent
expansion of a function without having to perform its full
reconstruction first.

The Laurent expansion algorithm is another algorithm whose node embeds
a subgraph.  Consider a graph $G$ representing a multi-valued
$(n+1)$-variate rational function in the variables $\{\epsilon,\z\}$.
The Laurent expansion node takes a list of length $n$ as input, which
represents the variables $\z$, and returns for each output of $G$ the
coefficients of its Laurent expansion in the first variable $\epsilon$,
up to a given order in $\epsilon$.

Without loss of generality, we only implement Laurent expansions
around $\epsilon=0$.  Expansions around other points, including
infinity, can be achieved by combining this node with another one
implementing a change of variables, which in turn can be represented
by an algorithm evaluating rational functions.

When the node is defined, we also specify the order at which we want
to truncate the expansion.  We can specify a different order for each
entry of the output of $G$.  This node has a learning phase, during
which it performs two univariate reconstructions in $\epsilon$ of the
output of $G$, for fixed numerical values of the variables $\z$.  The
first reconstruction uses Thiele's formula, and it is used to learn
the total degrees in $\epsilon$ of the numerators and the denominators
of the outputs of $G$.  Subsequent reconstructions will use the
univariate system-solving strategy discussed in
section~\ref{sec:mult-funct-reconstr}.  For each output of $G$, any
overall prefactor $\epsilon^p$, where $p$ can be a positive or
negative integer, is also detected and factored out to simplify
further reconstructions (notice that, after this, we can assume the
denominators to have the constant term equal to one).  These prefactors
also determine the starting order of the Laurent expansion, which
therefore is known after the learning phase.  The second
reconstruction in the learning phase is simply used as a consistency
check.

On each numerical evaluation, for given values of the inputs $\z$,
this node performs a full univariate reconstruction in $\epsilon$ of
the output of $G$ and then computes its Laurent expansion up to the
desired order.  Numerical evaluations of $G$ are cached so that they
can be reused for reconstructing several entries of its output for the
same values of $\z$.  The coefficients of the Laurent expansions of
each element are then chained together and returned.

\subsection{Algorithms with no input}
\label{sec:algorithms-with-no}

We finally point out that it is possible to define nodes and graphs
with no input.

Nodes with no input correspond to algorithms whose output may only
depend on the prime field $\Z_p$.  Some notable examples are nodes
implementing the solution of linear systems and linear fits (already
discussed in the previous sections) in the special case where they do
not depend on any list of free parameters $\z$.  Another example is a
node evaluating a list of rational numbers over a finite field
$\Z_p$.  Nodes with no input have depth zero, by definition.

A graph with no input is a graph with no input node.  The nodes with
the lowest depth of such a graph are nodes with no input.  The output of
this graph only depends on the prime field $\Z_p$ used.  These graphs
thus represent purely numerical (and rational) algorithms and no
functional reconstruction is therefore needed.  For these, we perform a
rational reconstruction of their output by combining Wang's algorithm
and the Chinese remainder theorem, as explained in
section~\ref{sec:finite-fields}.

\section{Reduction of scattering amplitudes}
\label{sec:reduct-scatt-ampl}

One of the most important and phenomenologically relevant applications
of the methods described in this paper is the reduction of scattering
amplitudes to a linear combination of master integrals or special
functions.  This is indeed a field which, in recent years, has
received a notable boost in our technical capabilities, thanks to the
usage of finite fields and functional reconstruction techniques.  In
particular, the results
in~\cite{Badger:2018enw,Henn:2019rmi,Badger:2019djh} have been
obtained using an in-development version of the framework presented
here.

\subsection{Integration-by-parts reduction to master integrals}
\label{sec:reduct-mast-integr}

Loop amplitudes are linear combinations of Feynman integrals.
Consider an $\ell$-loop amplitude $A$, or a contribution to it, with
$e$ external momenta $p_1,\ldots, p_{e}$.  The amplitude, in
dimensional regularization, is a linear combination of integrals over
the $d$-dimensional components of the loop momenta
$k_1,\ldots,k_\ell$.  It is convenient to write down these integrals
in a standard form.  For each topology $T$, let $\{D_{T,j}\}_{j=1}^n$
be a complete set of loop propagators, including auxiliary propagators
or irreducible scalar products, such that any scalar product of the
form $k_i\cdot k_j$ and $k_i\cdot p_j$ is a linear combination of
them.  In principle, there could also be scalar products of the form
$k_i\cdot \omega_j$ where $\omega_j$ are vectors orthogonal to the
external momenta $p_j$, but these can be integrated out in terms of
denominators $D_{T,j}$ an auxiliary (see e.g.\
ref.~\cite{Mastrolia:2016dhn}), hence they are not considered here.
Effective methods for obtaining this representation of an amplitude
are integrand reduction (discussed in
section~\ref{sec:integrand-reduction}) and the decomposition into form
factors (discussed in section~\ref{sec:tens-reduct-decomp}).  Hence,
given a list of integers $\vec{\alpha} = (\alpha_1,\ldots,\alpha_n)$,
we consider Feynman integrals with the standard form
\begin{align} \label{eq:canonicalint}
  I_{T,\vec{\alpha}}^{(d)} ={}& \int \left( \prod_j d k_j \right)\, \frac{1}{D_{T,1}^{\alpha_1}\cdots D_{T,n}^{\alpha_n}}.
\end{align}
Notice that the exponents $\alpha_j$ may be positive, zero, or negative.

Amplitudes may be written as linear combinations of the integrals
above as
\begin{equation}
  \label{eq:ampfunred}
  A = \sum_{j\in\{(T,\vec{\alpha})\}}\, a_j\, I_j,
\end{equation}
where the coefficients $a_j$ are rational functions of kinematic
invariants, and possibly of the dimensional regulator
$\epsilon = (4-d)/2$.  While the computation of the coefficients $a_j$
can be highly non-trivial for high-multiplicity processes, in this
section we assume them to be known.  Notice that they don't need to be
known analytically, but it is sufficient to have a numerical algorithm
for obtaining them.  As already mentioned, popular and successful
examples of these algorithms are integrand reduction and the
decomposition into form factors, which we will talk about in
sections~\ref{sec:integrand-reduction}
and~\ref{sec:tens-reduct-decomp}.

In general, the integrals $I_j$ appearing in Eq.~\eqref{eq:ampfunred}
are not all linearly independent.  Indeed they satisfy linear
relations such as integration-by-parts (IBP) identities, Lorentz
invariance identities, symmetries, and mappings.  The collection of
these relations form a large and sparse system of equations satisfied
by these integrals.  The most well known and widely used method for
generating such relations is the Laporta
algorithm~\cite{Laporta:2001dd}.  In this case, these identities can
be easily generated using popular computer algebra systems, especially
with the help of public tools (for instance, the package
\textsc{LiteRed}~\cite{Lee:2012cn} is very useful for generating these
relations in \textsc{Mathematica}).  However, any other method can be
used for building this system, as long as this is provided in the
form of a set of linear relations satisfied by Feynman integrals.

As explained in section~\ref{sec:dense-sparse-linear}, in order to
properly define this system we need to introduce an ordering between
the unknowns, in this case, the integrals $I_j=I_{T,\vec{\alpha}}$, by
assigning a weight to them~\cite{Laporta:2001dd}.  The efficiency of
the linear solver, as well as the number of equations left after
applying the mark-and-sweep method described in
section~\ref{sec:dense-sparse-linear}, strongly depends on this
ordering.  However, there is no unique good choice of it, and any
choice can be specified when the system is defined.  An example which
we found has good properties and prefers integrals with no higher
powers of denominators is provided in
Appendix~\ref{sec:observ-integr-parts}.

By solving this large system, which we henceforth refer to as
\emph{IBP system}, we reduce the amplitude to a linear combination of
a smaller set of integrals $G_j$, known as \emph{master integrals}
(MIs),
\begin{equation}
  \label{eq:ibpred}
  I_j = \sum_{k\in \textrm{MIs}} c_{jk}\, G_k,
\end{equation}
where the coefficients $c_{jk}$ are rational functions of the
kinematic invariants and the dimensional regulator $\epsilon$.  Notice
that the master integrals $G_k$ do not need to have the form in
Eq.~\eqref{eq:canonicalint}, but they can be arbitrary combinations of
integrals of that form.  In general, one may have a list of preferred
integrals which are defined as special linear combinations of those
in Eq.~\eqref{eq:canonicalint} characterized by good properties, such
as a simpler pole structure or a better analytic behaviour (a
convenient property to have is uniform transcendental
weight~\cite{Henn:2013pwa}, see also
section~\ref{sec:diff-equat-epsil}).  In such cases, we add the
definition of these integrals to the system of equations and we assign
to them a lower weight so that they are automatically chosen as
independent integrals, to the extent that this is possible, during the
Gauss elimination.  Another important fact to note is that the list
of master integrals is determined after the learning phase of the
linear solver, which only requires a few numerical evaluations.

After IBP reduction, amplitudes are written as linear combinations of
master integrals
\begin{equation}
  \label{eq:ampfred}
  A = \sum_{k\in \textrm{MIs}} A_k\, G_k,
\end{equation}
where the coefficients $A_k$, which are rational functions of
the kinematic invariants and the dimensional regulator $\epsilon$, can
be obtained via a matrix multiplication between the coefficients of
the unreduced amplitude in Eq.~\eqref{eq:ampfunred} and the ones in
the IBP solutions in Eq.~\eqref{eq:ibpred},
\begin{equation}
  \label{eq:5}
  A_k = \sum_{j} a_{j}\, c_{jk}.
\end{equation}

\begin{figure}[t]
  \centering
  \includegraphics[scale=0.35]{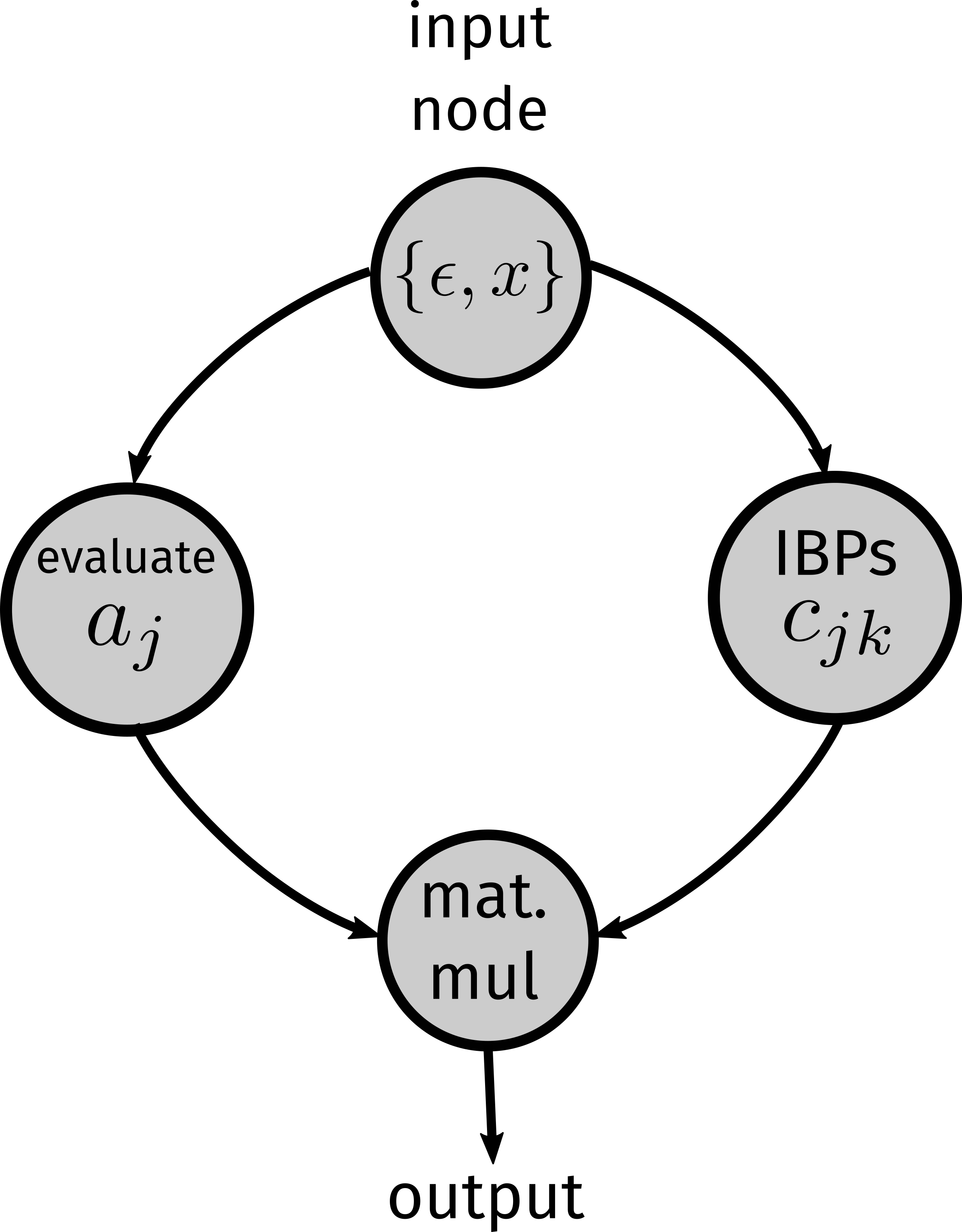}
  \hspace{1cm}
  \includegraphics[scale=0.46]{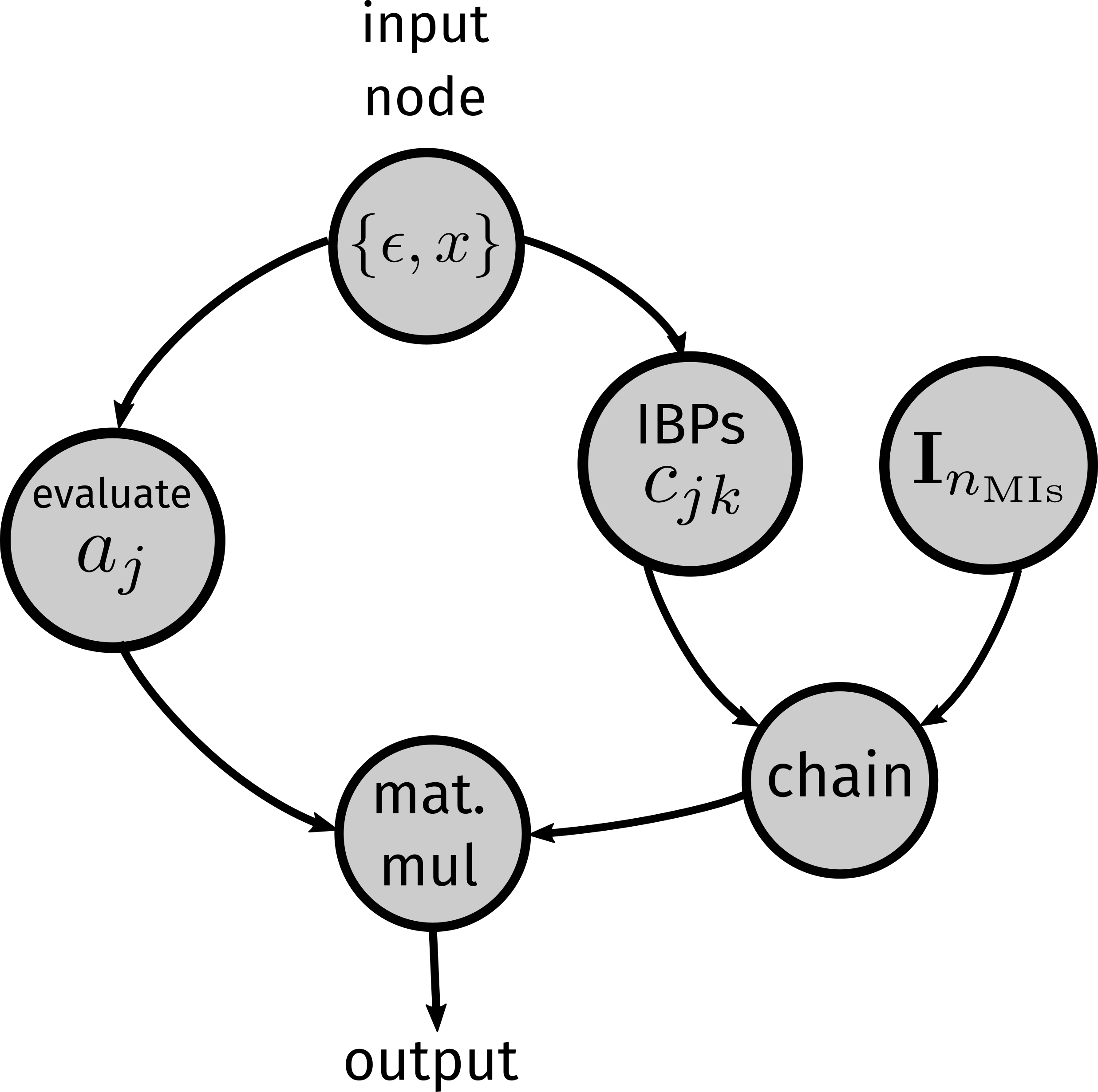}
  \caption{Two dataflow graphs representing the reduction of a
    scattering amplitude to master integrals.  The graph on the right
    has two additional nodes chaining to the coefficients of the IBP
    solutions an identity matrix, which represents the (trivial)
    reduction of the master integrals themselves.  These nodes are
    needed when the masters can also appear on the r.h.s.\ of the
    unreduced amplitude in Eq.~\eqref{eq:ampfunred}.  }
  \label{fig:amp}
\end{figure}

Putting these ingredients together, it is very easy to define a simple
dataflow graph representing this calculation, which is depicted in
fig.~\ref{fig:amp}.
\begin{itemize}
\item The input node of the graph represents the variables $\{\epsilon,x\}$ where
  $\epsilon$ is the dimensional regulator and $x$ can be any number of
  kinematic invariants.
\item The node $a_{j}$ takes as input the input node
  $\{\epsilon,x\}$ and evaluates the coefficients of the unreduced
  amplitude in Eq.~\eqref{eq:ampfunred}.  If these are known
  analytically this can simply be a node evaluating a list of rational
  functions, otherwise, it can represent something more complex, such as
  one of the algorithms we will discuss later.
\item The IBP node is a sparse linear solver which takes as input the
  input node $\{\epsilon,x\}$ and returns the coefficients $c_{jk}$
  obtained by numerically solving the IBP system.  Because these
  systems are homogeneous, we only return the homogeneous part of the
  solutions (the removed constant terms are zero).  After the
  learning phase is completed, we strongly recommend running the
  mark-and-sweep algorithm to reduce the number of equations.
\item Finally, the output node, which can be defined after the
  learning phase of the IBP node has been completed, is a matrix
  multiplication which takes as inputs the node $a_{j}$ and the IBP
  node.
\end{itemize}
The graph we just described, which is depicted on the left of
fig.~\ref{fig:amp}, ignores a technical subtlety.  The reduction
coefficients $c_{jk}$ returned by the IBP node express the non-master
integrals in terms of master integrals.  However, depending on our
choice of masters, the master integrals themselves may also appear on
the r.h.s.\ of the unreduced amplitude in Eq.~\eqref{eq:ampfunred}.
This creates a mismatch which does not allow to properly define the
final matrix multiplication.  More explicitly, if $n_{\textrm{MIs}}$
is the number of master integrals, and $n_{\textrm{non-MIs}}$ is the
number of non-master integrals appearing in Eq.~\eqref{eq:ampfunred},
then the IBP node returns a
$n_{\textrm{non-MIs}} \times n_{\textrm{MIs}}$ matrix.  However, if
the $n_{\textrm{MIs}}$ masters also appear on the r.h.s.\ of
Eq.~\eqref{eq:ampfunred}, then the output of the $a_j$ node has length
$n_{\textrm{non-MIs}} + n_{\textrm{MIs}}$, which makes it incompatible
with the IBP solution matrix it should be multiplied with.  This can,
however, be easily fixed by defining an additional node representing
the reduction of the master integrals to themselves, which is
trivially given by the $n_{\textrm{MIs}} \times n_{\textrm{MIs}}$
identity matrix $\mathbf{I}_{n_{\textrm{MIs}}}$ (this is a node with
no input, which evaluates a list of rational numbers, see also
section~\ref{sec:algorithms-with-no}).  After this is chained (see
section~\ref{sec:basic-oper-lists}) to the output of the IBP node, we
obtain a
$(n_{\textrm{non-MIs}}+n_{\textrm{MIs}}) \times n_{\textrm{MIs}}$
matrix containing the reduction to master integrals of all the
$n_{\textrm{non-MIs}} + n_{\textrm{MIs}}$ Feynman integrals in
Eq.~\eqref{eq:ampfunred}.  Hence the final matrix multiplication is
well defined.  This graph is depicted on the right of
fig.~\ref{fig:amp}.  Notice that these two extra nodes are not
necessary when all the master integrals have been separately defined
and don't appear in our representation of the unreduced amplitude,
because in this case the output of the $a_j$ node has length
$n_{\textrm{non-MIs}}$ and can be directly multiplied with the matrix
computed by the IBP node.

The dataflow graph we just described computes the coefficients of the
reduction of an amplitude to master integrals.  By evaluating this
graph several times, one can thus reconstruct the analytic expressions
of these coefficients, without the need of deriving large and complex
IBP tables.  This represents a major advantage, since IBP tables for
complex processes can be extremely large, significantly more complex
than the final result for the reduced amplitude, hard to compute, and
also hard to use -- since they require to apply a huge list of complex
substitutions to the unreduced amplitude.  On the other hand, using
the approach described here, IBP tables are always computed
numerically, and only the final result is reconstructed analytically.
Hence, by building a very simple dataflow graph consisting of only a
few nodes, we are able to sidestep the bottleneck of computing and
using large, analytic IBP tables.  This approach has already allowed
(e.g.\ in ref.s~\cite{Badger:2018enw,Badger:2019djh}) to perform
reductions in cases where the IBP tables are known to be too large and
complex to be computed and used with reasonable computing resources.

\subsection{Reduction to special functions and Laurent expansion in
  $\epsilon$}
\label{sec:reduct-spec-funct}

The expansion in the dimensional regulator $\epsilon$ of the master
integrals can often be computed in terms of special functions, such as
multiple polylogarithms or their elliptic generalization.  When this
is possible, the result for the $\epsilon$ expansion of a scattering
amplitude might be significantly simpler than the one in terms of
master integrals.  For the sake of argument, we assume to be
interested in the poles and the finite part of the amplitude, but
everything we are going to discuss can be easily adapted to different
requirements.

Let $\{f_k=f_k(x)\}$ be a complete list of special functions (which
may also include numerical constants) such that every master integral
$G_j$, expanded up to its finite part, can be expressed in terms of
these as
\begin{equation}
  \label{eq:mis2fs}
  G_j = \sum_{jk}\, g_{jk}(\epsilon,x)\, f_k + \O(\epsilon),
\end{equation}
where $g_{jk}$ are rational functions in $\epsilon$ and $x$
(typically, they will be a Laurent polynomial in $\epsilon$, but this
is not important for the discussion).  Recalling
Eq.~\eqref{eq:ampfred}, we can thus write the amplitude in terms of
these functions as
\begin{equation}
  \label{eq:7}
  A = \sum_k\, u_k(\epsilon,x)\, f_k + O(\epsilon),
\end{equation}
where the rational functions $u_k$ are defined as
\begin{equation}
  \label{eq:ampfucoeffs}
  u_k(\epsilon,x)= \sum_j\, A_j(\epsilon,x)\, g_{jk}(\epsilon,x).
\end{equation}
We are interested in the expansion in $\epsilon$ of the coefficients
$u_k$, i.e.\ in the coefficients $u_{k}^{(j)}=u_{k}^{(j)}(x)$ such that
\begin{equation}
  \label{eq:ampulaurexpansion}
  u_{k}(\epsilon,x) = \sum_{j=-p}^0 u_{k}^{(j)}(x)\, \epsilon^j  + \O(\epsilon),
\end{equation}
where $p$ is such that the leading pole of the amplitude is
proportional to $\epsilon^{-p}$.

\begin{figure}[t]
  \centering
  \includegraphics[scale=0.35]{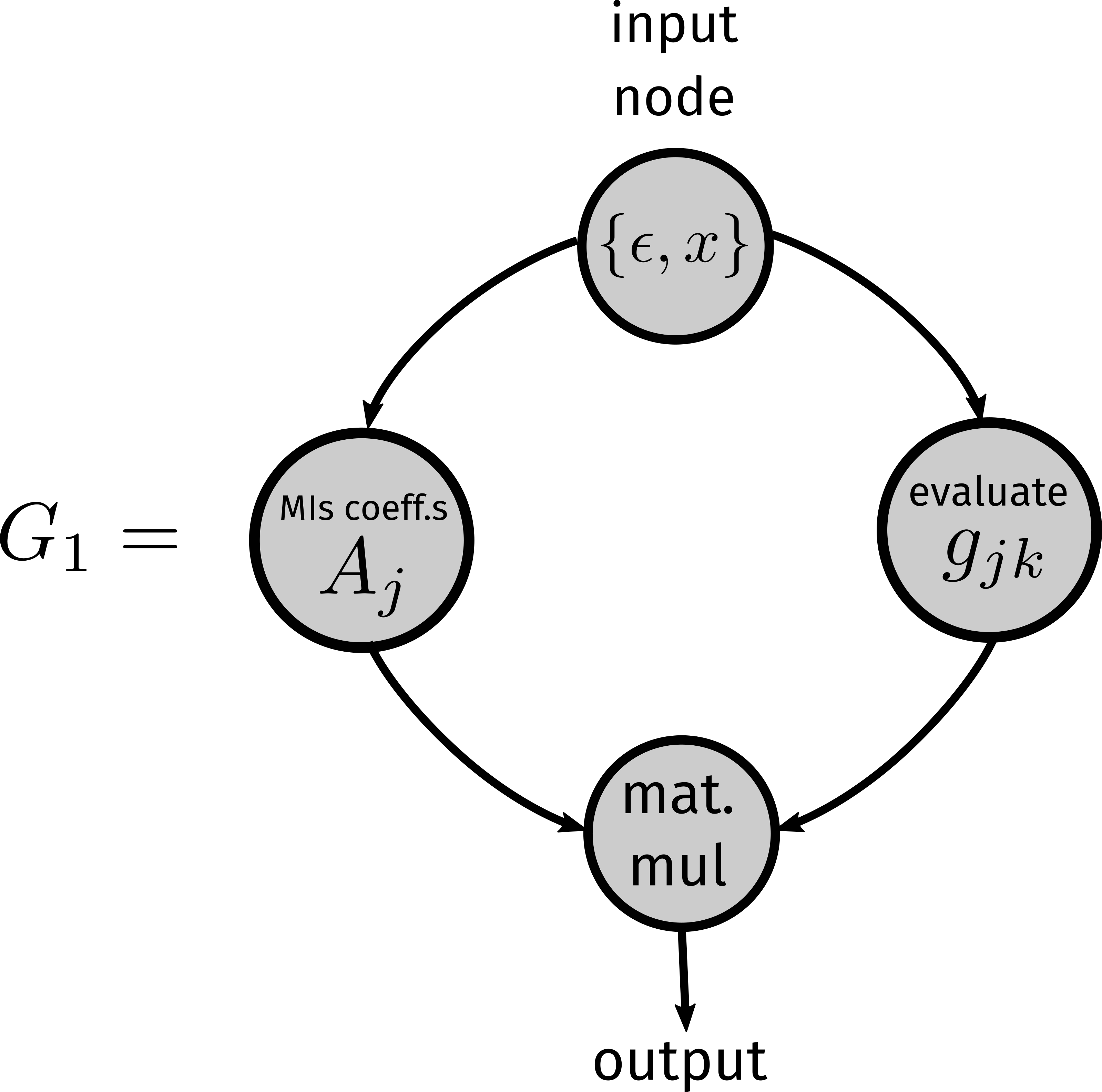}
  \hspace{2cm}
  \includegraphics[scale=0.35]{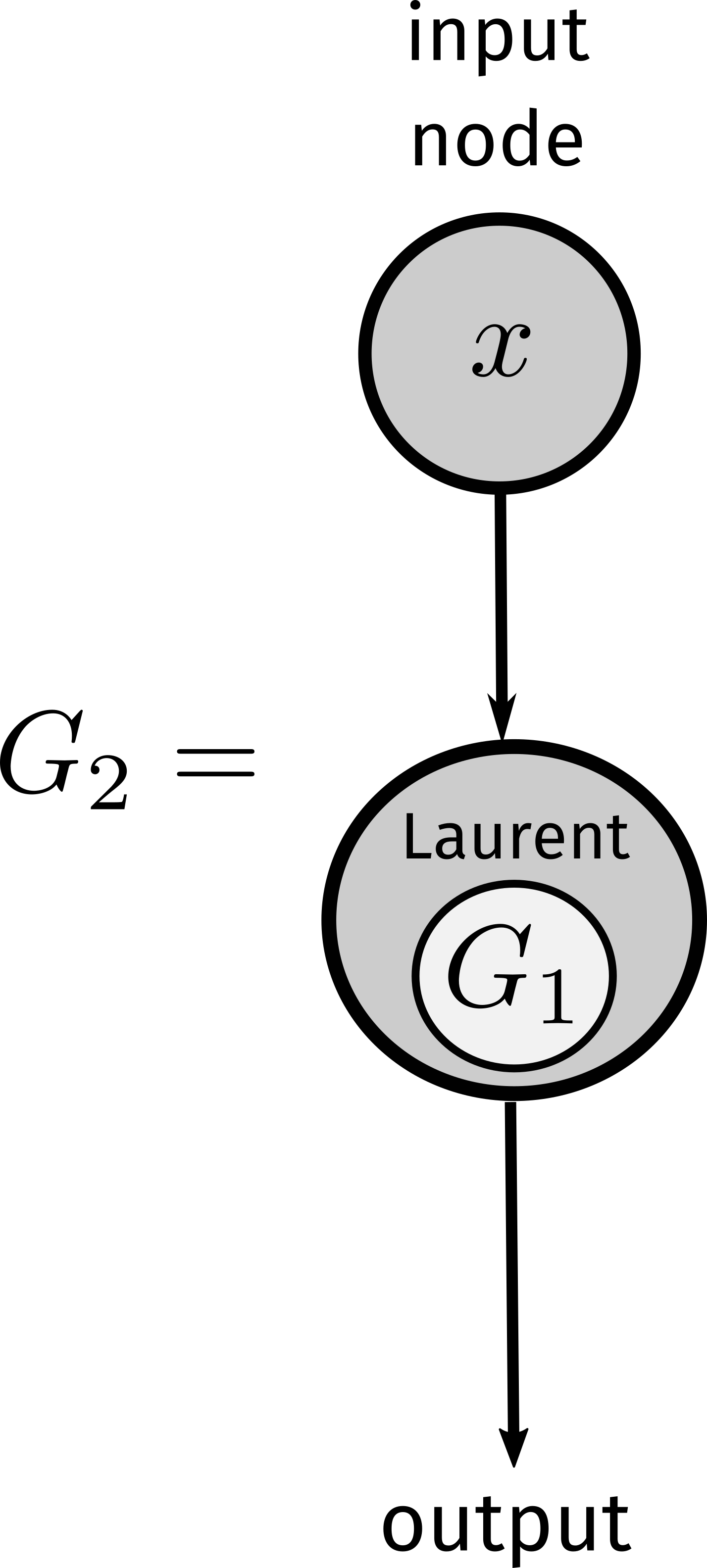}
  \caption{Two graphs which, combined, compute the $\epsilon$
    expansion of the coefficients of scattering amplitudes in terms of
    special functions.  In the first graph $G_1$, $A_{j}$ represents
    the calculation of the coefficients of the master integrals
    presented in section~\ref{sec:reduct-mast-integr} and
    fig.~\ref{fig:amp}.  The graph $G_2$ then takes the graph $G_1$ as
    subgraph in one of its nodes, which computes its Laurent expansion
    in $\epsilon$.}
  \label{fig:amp2f}
\end{figure}

Computing the coefficients $u_{k}^{(j)}(x)$ in our framework is
straightforward.  We start from the dataflow graph described in
section~\ref{sec:reduct-mast-integr}, which computes the coefficients
$A_j$ of the master integrals.  We first extend this graph in order to
get the unexpanded coefficients $u_{k}(\epsilon,x)$.  This is simply
done by adding a node $g_{jk}$, which evaluates the rational functions
$g_{jk}(\epsilon,x)$ defined in Eq.~\eqref{eq:mis2fs}, and a matrix
multiplication node between the node $A_{j}$ (which was the output
node in the previous case) and $g_{jk}$, as one can see from
Eq.~\eqref{eq:ampfucoeffs}.  Let us call this dataflow graph $G_1$.
We then create a new graph $G_2$ with input variables $x$.  Inside the
latter, we create a Laurent expansion node, which takes as its subgraph
$G_1$.  The output of this node will be the coefficients $u_{k}^{(j)}$
of the Laurent expansion in Eq.~\eqref{eq:ampulaurexpansion}.  This is
depicted in fig.~\ref{fig:amp2f}.

Because the coefficients $u_{k}^{(j)}(x)$ might not be all linearly
independent, we also recommend running the \emph{subgraph fit}
algorithm described in section~\ref{sec:linear-fit} in order to find
linear relations between them.  In particular, this can be used to
rewrite the most complex coefficients as linear combinations of
simpler ones, yielding thus a more compact form of the result, which
is also easier to reconstruct.

We finally point out that one can further elaborate the graph $G_1$ in
order to include \emph{renormalization}, \emph{subtraction} of
infrared poles, and more.  This is done by rewriting these
subtractions, which are typically known analytically since they depend
on lower-loop results, in terms of the same list of functions
$\{f_k\}$ as the amplitude.  After doing so, the coefficients of the
subtraction terms multiplying the functions $f_k$ are added to the
graph as nodes evaluating rational functions and summed to the
output using the Addition node described in
section~\ref{sec:basic-oper-lists}.  This may thus simplify the output
of the Laurent expansion computed in the graph $G_2$, which will,
therefore, be easier to reconstruct.

It goes without saying that, even if we focused on scattering
amplitudes, the same strategy can be applied to other objects in
quantum field theory which have similar properties, such as
correlation functions and form factors.

\section{Differential equations for master integrals}
\label{sec:diff-equat-mast}

Integration-by-parts identities are not only useful to reduce
amplitudes to linear combinations of a minimal set of independent
master integrals, but they are also helpful for the calculation of the
master integrals themselves via the method of differential
equations~\cite{Kotikov:1990kg,Gehrmann:1999as}.  Indeed the master
integrals $G_j$ satisfy systems of coupled partial differential
equations with respect to the invariants $x$,
\begin{equation}
  \label{eq:de}
  \frac{\partial}{\partial x}\, G_j = \sum A_{jk}^{(x)}(\epsilon,x)\, G_k.
\end{equation}
Solving these systems of differential equations is one of the most
effective and successful methods for computing the master integrals.

\subsection{Reconstructing differential equations}
\label{sec:reconstr-diff-equat}

The differential equation matrices can be easily computed within our
framework, using a strategy which is completely analogous to the one
described in section~\ref{sec:reduct-mast-integr} for the reduction of
scattering amplitudes to master integrals.

We first determine the master integrals by solving the IBP system
numerically over finite fields.  For this, we need to specify a list
of \emph{needed integrals}, i.e.\ a list of needed unknowns for which
the system solver is asked to provide a solution since in general one
cannot reduce to master integrals all the integrals appearing in an
IBP system.  We then make a conservative choice which is likely to be
a superset of all the integrals which need to be reduced for computing
the differential equations.

Then, the derivatives of master integrals with respect to kinematic
invariants can be easily computed analytically,
\begin{equation}
  \label{eq:deunreduced}
  \frac{\partial}{\partial x}\, G_j = \sum_{k\in (T,\vec{a})} a^{(x)}_{jk} \, I_k,
\end{equation}
where the integrals $I_j$ have the standard form defined in
Eq.~\eqref{eq:canonicalint}, and $a^{(x)}_{jk}$ are rational functions
of the invariants $x$.  At this stage, we may reset the list of needed
unknowns of the IBP system to include only the ones appearing on the
r.h.s.\ of Eq.~\eqref{eq:deunreduced}.  After that, we also strongly
suggest running the mark-and-sweep algorithm for removing unneeded
equations.

By solving the IBP system, we reduce the integrals $I_j$ to master
integrals.  This defines the coefficients $c_{jk}$ of the reduction,
as in Eq.~\eqref{eq:ibpred}.  The differential equation matrices
$A_{jk}^{(x)}$ are thus obtained via the matrix multiplication
\begin{equation}
  \label{eq:dematmul}
  A_{jk}^{(x)} = \sum_l a^{(x)}_{jl}\, c_{lk}.
\end{equation}

\begin{figure}[t]
  \centering
  \includegraphics[scale=0.35]{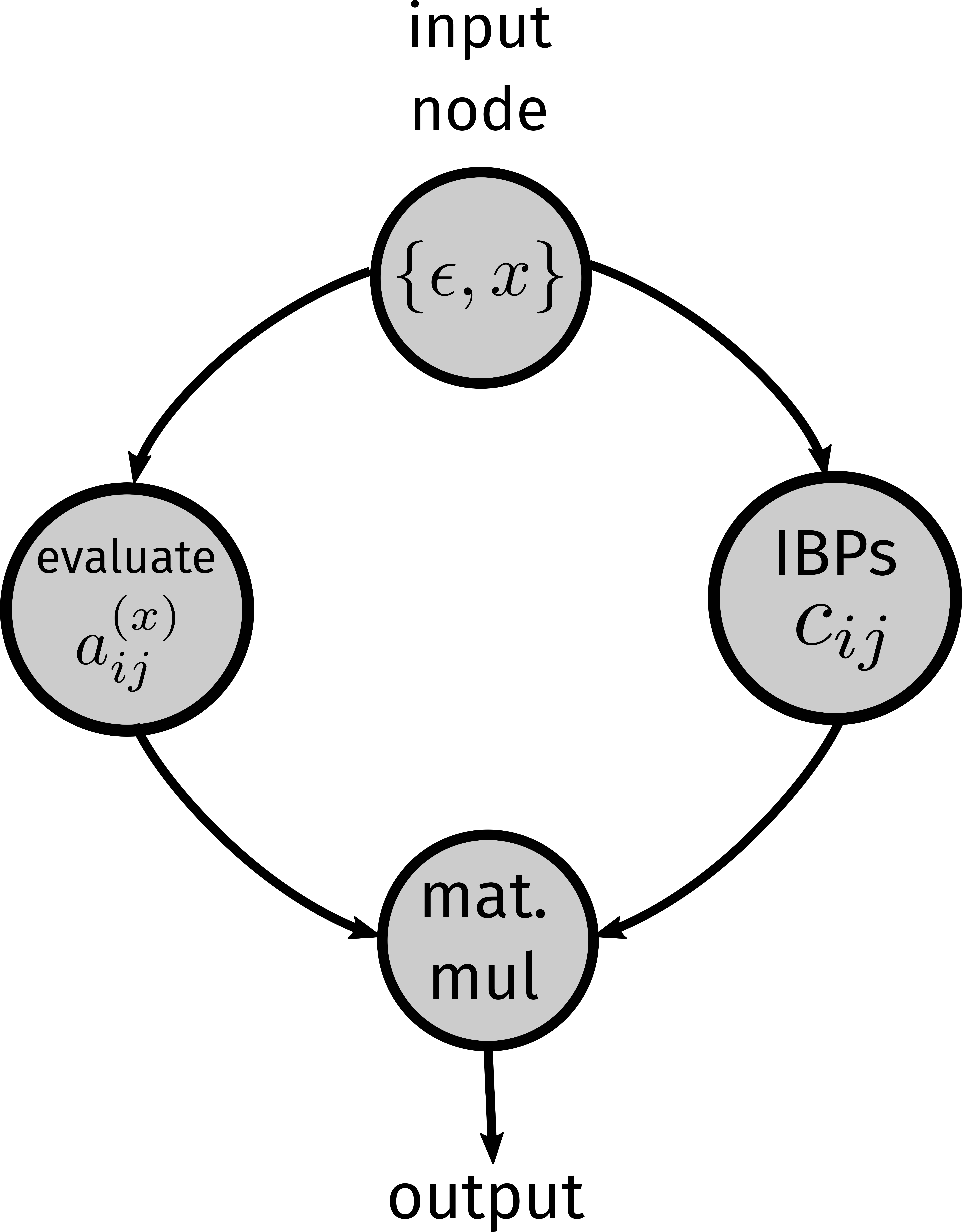}
  \hspace{2cm}
  \includegraphics[scale=0.35]{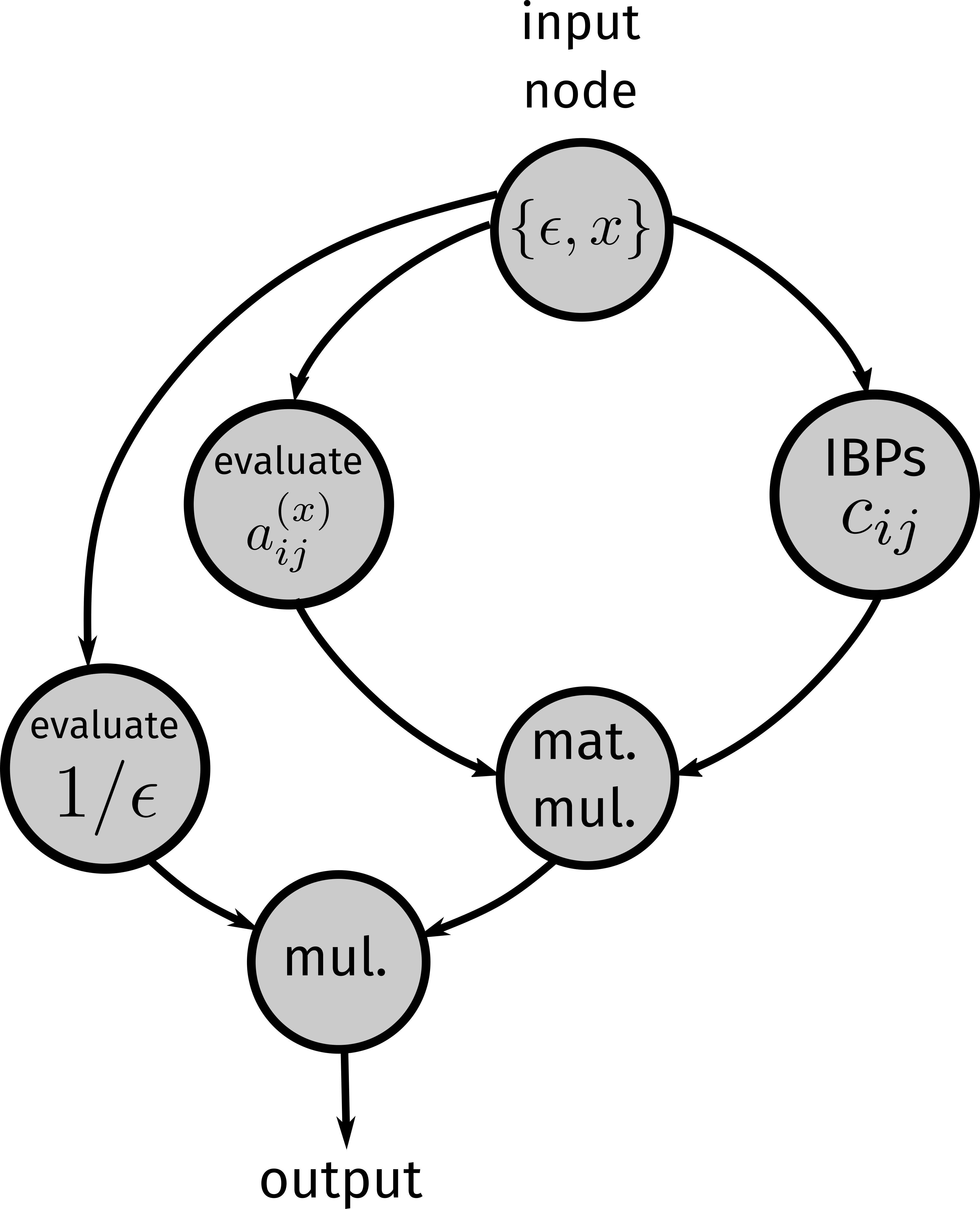}
  \caption{On the left, a dataflow graph representing the calculation
    of differential equations satisfied by master integrals.  It is
    similar to the one depicted in fig.~\ref{fig:amp} for the
    reduction of amplitudes to master integrals.  On the right, a
    dataflow graph computing the differential equation matrices
    divided by $\epsilon$.  As explained in
    section~\ref{sec:diff-equat-epsil} we can verify the
    $\epsilon$-form of the differential equations by checking
    numerically that the output of the latter graph does not depend on
    $\epsilon$.}
  \label{fig:de}
\end{figure}

A dataflow graph representing this calculation can, therefore, be almost
identical to the one described in
section~\ref{sec:reduct-mast-integr}, and it is depicted on the left
side of fig.~\ref{fig:de}.
In particular, it has an input node representing the variables
$\{\epsilon,x\}$, a node evaluating the rational functions
$a^{(x)}_{ij}$ appearing in the unreduced derivatives of
Eq.~\eqref{eq:deunreduced}, a node with the IBP system, and an output
node with the final matrix multiplication in Eq.~\eqref{eq:dematmul}.
Similarly to the case of the amplitudes, if the master integrals are
chosen such that they can also appear on the r.h.s.\ of the unreduced
derivatives in Eq.~\eqref{eq:deunreduced}, then we also add an
identity matrix node, and a node chaining this to the IBP node (see
section~\ref{sec:reduct-mast-integr} and fig.~\ref{fig:amp} for more
details).

By defining this graph, we can reconstruct the differential equations
of the master integrals directly, without the need of computing IBP
tables analytically, similarly to the case of the reduction of
amplitudes.  This usually yields a substantial simplification of the
calculation.

\subsection{Differential equations in $\epsilon$-form}
\label{sec:diff-equat-epsil}

It has been observed in ref.~\cite{Henn:2013pwa} that the differential
equation method becomes more powerful and effective if the master
integrals are chosen such that they are \emph{pure functions of
  uniform transcendental weight}, henceforth \emph{UT functions} for
brevity (we refer to ref.~\cite{Henn:2013pwa} for a definition).
Remarkably, as pointed out in ref.~\cite{Henn:2013pwa}, one can build
a list of integrals having this property without doing any reduction
at all, by using some effective rules or by analyzing the leading
singularities of Feynman integrals.  A systematic algorithm which
implements this analysis of leading singularities was developed and
described in~\cite{WasserMSc}, and recently extended in
ref.~\cite{Chicherin:2018old}.  Once a (possibly over-complete) list
of UT integrals has been found, their definitions can be added as
additional equations to the IBP system.  By assigning a lower weight
to these integrals, they will be automatically chosen as preferred
master integrals by the system solver.

If $\{G_k\}$ represents a basis of UT master integrals, the
differential equation matrices take the form~\cite{Henn:2013pwa}
\begin{equation}
  \label{eq:8}
  A^{(x)}_{ij}(\epsilon,x) = \epsilon\, A^{(x)}_{ij}(x),
\end{equation}
i.e.\ their $\epsilon$ dependence is simply an $\epsilon$ prefactor.
This greatly simplifies the process of solving the system
perturbatively in $\epsilon$.  When this happens, the system of
differential equations is said to be in $\epsilon$-form or in
canonical form.

If a list of UT candidates is known, as we said, we may add their
definition to the IBP system, and then we divide the final result for
the matrices by $\epsilon$.  This is done by modifying the dataflow
graph defined before, with the addition of a node which evaluates the
rational function $1/\epsilon$, and a new output node which multiplies
the $1/\epsilon$ node with the older output node.  As mentioned in
section~\ref{sec:basic-oper-lists}, the multiplication can be
accomplished using a matrix multiplication node, which interprets
$1/\epsilon$ as a $1\times 1$ matrix and its second input node as a
matrix with only one row.  This modified graph is depicted on the
right side of fig.~\ref{fig:de}.  Once the graph is defined, we can
evaluate it numerically for several values of $\epsilon$ while keeping
$x$ fixed, in order to check that the system is indeed in
$\epsilon$-form.

Differential systems in $\epsilon$-form for UT integrals are typically
much easier to reconstruct since they have a particularly simple
functional structure.  Hence they benefit even more from the
functional reconstruction methods described in this paper, which allow
to reconstruct this result directly without dealing with the
significantly more complex analytic intermediate expressions one would
have in a traditional calculation.

A large class of Feynman integrals can be written as linear
combinations of iterated integrals of the form (using the notation
in~\cite{Goncharov:2010jf})
\begin{equation}
  \label{eq:dlogiterated}
  \int d \log w_1 \circ d \log w_2 \circ \cdots \circ d \log w_n
\end{equation}
where the $d \log$ arguments $w_k$ are commonly called \emph{letters}.
A complete set of letters is called \emph{alphabet}.  While the
alphabet of a multi-loop topology is often inferred from the
differential equations for the master integrals, there are some cases
where this can instead be guessed a priori.  In such cases, finding
differential equations for UT master integrals can be even simpler,
since the calculation can be reduced to a numerical linear
fit~\cite{Abreu:2018rcw}.  Indeed, it is well known that differential
equations matrices for UT master integrals, aside from their
$\epsilon$ prefactor, are expected to be linear combinations of first
derivatives of logarithms of letters, with rational numerical
coefficients.  More explicitly, if $W=\{w_1,w_2,\ldots\}$, with
$w_k=w_k(x)$, is the alphabet of a topology, the differential equation
matrices for a set of UT master integrals take the form
\begin{equation}
  A^{(x)}_{ij}(x) = \sum_k \frac{\partial \log (w_k)}{\partial x}\, C^{(x,k)}_{ij},
\end{equation}
where $C^{(x,k)}_{ij}$ are rational numbers.  Hence, rather than
employing multivariate functional reconstruction methods, in this case,
we can compute the differential equation matrices just with a linear
fit.  For this purpose, we can apply the \emph{subgraph multi-fit}
algorithm described in section~\ref{sec:linear-fit}.  More explicitly,
we create a graph $G_1$ whose output contains both the derivatives
$\partial \log w_k/\partial x$ of the letters and the (non-vanishing)
matrix elements $A^{(x)}_{ij}$.  We then build a second graph $G_2$,
with a subgraph multi-fit node containing $G_1$, which performs a fit
of each matrix element with respect to the basis of functions
$\{\partial \log w_k/\partial x\}$, as described in
section~\ref{sec:linear-fit}.  Notice that $G_2$ has no input node,
and therefore we run a numerical reconstruction of its output over
$\Q$ using Wang's algorithm and the Chinese remainder theorem, as
already explained in section~\ref{sec:algorithms-with-no}.

\subsection{Differential equations with square roots}
\label{sec:diff-equat-with}

In our discussion of differential equations for UT integrals, we have
so far neglected the potential issue of the presence of square roots
in their definition.  Indeed, there are cases where, in order to
define UT integrals, one needs to take rational linear combinations of
integrals of the form of Eq.~\eqref{eq:canonicalint} and multiply them
by a prefactor equal to the square root of a rational function of the
invariants $x$.  Even in cases where these square roots may be removed
via a suitable change of variables, one may still wish to compute
differential equations in terms of the original kinematic invariants,
at least as a first step.  While square roots may be accommodated in
our framework by considering finite fields which are more general than
$\Z_p$, we would like to point out in this section that this is not
necessary for computing differential equations.

Let us rewrite the master integrals $G_j$ as
\begin{equation}
  \label{eq:9}
  G_j = R_j\, G^{\textrm{r.f.}}_j,
\end{equation}
where $R_j$ is either equal to one or to the square root of a rational
function of the invariants $x$, and $\{G^{\textrm{r.f.}}_j\}$ are a
set of root-free master integrals, which can be written as rational
linear combinations of standard Feynman integrals of the form of
Eq.~\eqref{eq:canonicalint}.  We first observe that the quantity
\begin{equation}
  \label{eq:dederivrat}
  \frac{1}{R_j}\, \frac{\partial}{\partial x}\, G_j = \left( \frac{1}{R_j} \frac{\partial R_j}{\partial x} \right)\, G^{\textrm{r.f.}}_j + \frac{\partial}{\partial x}\, G^{\textrm{r.f.}}_j,
\end{equation}
which can be easily computed analytically, is also a rational
combination of standard Feynman integrals.  This is indeed manifest on
the r.h.s.\ of the equation, since if $R$ is the square root of a
rational function then $R'/R$ is rational.  This implies that, via IBP
identities, we can reduce the root-free quantity in
Eq.~\eqref{eq:dederivrat} to the root-free master integrals and obtain
\begin{equation}
  \label{eq:12}
  \frac{1}{R_j}\, \frac{\partial}{\partial x}\, G_j = \sum_k \tilde{A}^{(x)}_{jk}\, G^{\textrm{r.f.}}_k,
\end{equation}
where the matrix $\tilde{A}^{(x)}_{jk}$ is also rational since the
IBP reduction itself cannot introduce any non-rational factor.  One
can finally show that the matrix $\tilde{A}^{(x)}_{jk}$ is related to
the differential equation matrix $A^{(x)}_{jk}$ we wish to compute by
\begin{equation}
  \label{eq:2}
  A^{(x)}_{jk} = \frac{R_j}{R_k}\, \tilde{A}^{(x)}_{jk},
\end{equation}
i.e.\ simply by rescaling each matrix element by a prefactor.  We can,
therefore, apply the methods described above to the quantity in
Eq.~\eqref{eq:dederivrat} (rather than to the simple derivatives of
the master integrals), use it to reconstruct the rational matrix
$\tilde{A}^{(x)}_{jk}$, and finally recover $A^{(x)}_{jk}$ by
introducing the appropriate prefactors.  Notice also that
$A^{(x)}_{jk}$ is in $\epsilon$-form if and only if
$\tilde{A}^{(x)}_{jk}$ is in $\epsilon$-form.

\section{Integrand reduction}
\label{sec:integrand-reduction}

In section~\ref{sec:reduct-scatt-ampl}, we explained how to compute
the reduction of a scattering amplitude, either to a linear
combination of master integrals or to a combination of special
functions expanded in the dimensional regulator.  One of the
ingredients of the algorithm discussed there was a representation of
the unreduced amplitude (cfr.\ with Eq.~\eqref{eq:ampfunred}) as a
linear combination of Feynman integrals cast in a standard form, such
as the one in Eq.~\eqref{eq:canonicalint}.  In particular, within the
\textsc{FiniteFlow} framework, we need a numerical algorithm capable of
computing the coefficients $a_j$ of such a linear combination.  This
is trivial if an analytic expression is known for the $a_j$.  However,
this is not always the case.  Indeed, for complex processes, casting
the amplitude in such a form is a very challenging problem.  In this
section, we discuss \emph{integrand reduction}
methods~\cite{Ossola:2006us,Giele:2008ve,Mastrolia:2011pr,Badger:2012dp,Zhang:2012ce,Mastrolia:2012an,Mastrolia:2016dhn},
which are an efficient way of obtaining this representation of the
amplitude and are suitable for complex processes.

\subsection{Integrand reduction via linear fits}
\label{sec:integr-reduct-via}

Amplitudes are linear combinations of integrals of the form
\begin{equation}
  \label{eq:ampint}
    A ={} \int \left( \prod_j d k_j \right)\, \frac{\N(k_1,\ldots,k_\ell)}{D_{1}^{\alpha_1}\cdots D_{n}^{\alpha_n}}, \qquad \alpha_j>0,
  \end{equation}
  where $\N$ is a polynomial numerator in the loop components, and
  $D_j$ are denominators of loop propagators.  For simplicity, we
  consider only one topology, identified by a set of loop
  denominators, but we understand that the approach discussed here
  should be applied to all the topologies contributing to the
  amplitude we wish to compute.

  Integrand reduction methods rewrite the integrand as a linear
  combination of functions belonging to an \emph{integrand basis}
\begin{equation}
  \label{eq:intdec}
  \frac{\N(k_j)}{D_{1}^{\alpha_1}\cdots D_{n}^{\alpha_n}} = \sum_{\beta_j | 0\leq \beta_j\leq\alpha_j}\, \frac{\Delta_{\beta_1\cdots \beta_n}}{D_{1}^{\beta_1}\cdots D_{n}^{\beta_n}},
\end{equation}
where $\Delta_{\vec{\beta}} \equiv \Delta_{\beta_1\cdots \beta_n}$ has
the form
\begin{equation}
  \label{eq:delta}
\Delta_{\vec{\beta}}  = \sum_j c_{\vec{\beta},j}\, m_{\vec{\beta},j}(k_1,\ldots,k_\ell).
\end{equation}
In the previous equations, the functions $m_{\vec{\beta},j}$ are a
complete set of \emph{irreducible numerators}, i.e.\ numerators which,
at the integrand level, cannot be written in terms of the loop
propagators they are sitting on.  In other words, the terms
\begin{equation}
  \label{eq:intbasiselem}
  \frac{m_{\vec{\beta},j}(k_1,\ldots,k_\ell)}{D_{1}^{\beta_1}\cdots D_{n}^{\beta_n}} \qquad \textrm{with
  } 0\leq \beta_1\leq \alpha_1,\ldots, 0\leq \beta_n\leq \alpha_n,
\end{equation}
must form a complete basis of rational functions in the loop
components, for the loop integrand we are interested in.  The
coefficients $c_{\vec{\beta},j}$, which do not depend on the loop
momenta but only on the kinematic invariants, are unknowns which
parametrize an integrand in terms of the chosen basis.  The functions
$\Delta_{\vec{\beta}}$, also known in the literature as
\emph{residues} or \emph{on-shell integrands}, collect the elements of
the integrand basis which share the same loop-denominator structure.
The integrand basis can be chosen a priori solely based on the loop
topology, and independently of the process or the particle content of
the loop diagrams (see below for a few examples).  The parametric
coefficients $c_{\vec{\beta},j}$ in Eq.~\eqref{eq:delta} are instead
process dependent and represent the unknowns of this representation.

Once an integrand basis has been chosen, the unknown coefficients
$c_{\vec{\beta},j}$ can be determined via a \emph{linear fit}.  For
this purpose, we can use the algorithm described in
section~\ref{sec:linear-fit}, using kinematic invariants as the free
parameters $\z$, loop variables as the additional set of variables
$\mathbf{\tau}$, and $c_{\vec{\beta},j}$ as the unknowns of the
system.  In particular, in a dimensional regularization scheme where
the external states are four-dimensional (such as the
t'Hooft-Veltman\cite{tHooft:1972fi} and
Four-Dimensional-Helicity~\cite{Bern:2002zk} schemes), the integrand
depends on
\begin{equation*}
  4\, \ell + \frac{\ell (\ell+1)}{2}
\end{equation*}
loop variables.  These can be chosen, for instance, as the
four-dimensional components of the loop momenta with respect to a
basis of four-dimensional vectors, plus the independent scalar
products between the extra-dimensional projections of the loop momenta
\begin{equation}
  \label{eq:muij}
  \mu_{ij} = -k_i^{[-2\epsilon]}\cdot k_j^{[-2\epsilon]}.
\end{equation}
While performing a global fit of all the coefficients at the same time
is theoretically possible, in practice it is extremely inefficient and
impractical, because it involves solving a dense system of linear
equations of the same size as the number of the unknown coefficients.
One can however greatly simplify the problem by splitting it into
several smaller linear fits, using the so-call \emph{fit-on-the-cut}
approach~\cite{Ossola:2006us}.  This consists of evaluating the
integrand on \emph{multiple cuts}, i.e.\ values of the loop momenta
such that a subset of loop propagators vanish (we also understand that
vanishing denominators should be removed from the integrand when
applying a cut).  On each cut, we also have fewer independent loop
variables $\mathbf{\tau}$, namely those which are not fixed by the cut
conditions.  This method is best used in a top-down approach.  We
first cut (i.e.\ set to zero) as many propagators as possible, and use
linear fits on maximal cuts for determining a first set of
coefficients.  We then proceed with linear fits on cuts involving
fewer and fewer propagators.  When performing a fit on a multiple cut,
on-shell integrands which have already been fixed on previous cuts are
first subtracted from the integrand.  These subtractions are sometimes
referred to as \emph{subtractions at the integrand level}.  If an
integrand has all denominator powers $\alpha_j$ equal to one, with
this approach we determine the coefficients of one and only one
on-shell integrand $\Delta_{\vec{\beta}}$ on each cut.  If higher
powers of propagators are present, more than one off-shell integrand
must be determined at the same time on some cuts, but this doesn't
qualitatively change the algorithm for the linear fit (this point is
discussed more in detail in
ref.s~\cite{Abreu:2017idw,Badger:2017jhb}).

Subtractions at the integrand level can be implemented using the
linear fit algorithm described in Eq.~\eqref{eq:linfitaux}.  In
particular, we define a dataflow graph where each multiple cut
corresponds to a different node, whose output is the list of coefficients
$c_{\vec{\beta},j}$ determined by a linear fit.  Each node takes as
input, besides the kinematic variables $\z$, the output of all the
higher-point cuts with non-vanishing subtractions on the current cut.
The coefficients returned by the input nodes will be used as weights
$w_j$ (cfr.\ with Eq.~\eqref{eq:linfitaux}) for the subtractions,
while the integrand will typically have weight one.  Notice that the
linear fit described in Eq.~\eqref{eq:linfitaux} also allows to define
a set of auxiliary functions, in terms of which we can express both
the integrand and the integrand basis.  This is very convenient since
it allows to express these objects in terms of scalar products, spinor
chains, or other auxiliary functions which may yield a simple
representation.  Hence, we only have to explicitly substitute the cut
solutions inside these functions, which are then evaluated
numerically.  In particular, we don't need to substitute the cut
solutions inside the full integrand or the full set of integrand basis
elements appearing in the subtraction terms, which may yield
complicated expressions in some cases.

We also note that, when using the loop variables described above,
finding a rational parametrization of the cut solutions is a simple
problem of linear algebra.  As already explained
in~\cite{Mastrolia:2016dhn} one can proceed by splitting the cut
denominators into categories, such that denominators in the same
category depend on the same subset of loop momenta.  For each
category, we choose a representative, and we take differences between
all the other denominators and this representative.  This gives a
linear system of equations for the four-dimensional components of the
loop momenta which live in the space spanned by the external legs.
Next, we complete this solution by setting to zero the representatives
of each category.  This gives a system of equations which is linear in
the variables $\mu_{ij}$.  Notice that this is only true when we work
in $d$ dimensions.

If neither the integrand nor the integrand basis depends on the
dimensional regulator $\epsilon$, it is convenient to embed the
integrand reduction nodes in a \emph{memoized subgraph}, as described
at the end of section~\ref{sec:subgraphs}.  During the Laurent
expansion, this avoids repeating the integrand reduction for several
values of $\epsilon$ and fixed values of the kinematic invariants.  If
the integrand has a polynomial dependence on $\epsilon$, as it happens
for amplitudes in the t’Hooft-Veltman regularization scheme, we can
still implement this improvement by using several memoized subgraphs,
i.e.\ one for each power of $\epsilon$ in the numerator.

The algorithm we described allows to define a dataflow graph
implementing a full multi-loop integrand reduction over finite fields,
starting from a known integrand and an integrand basis.  This is
particularly convenient when using \textsc{FiniteFlow} from a computer
algebra system.  The output of all these nodes can then be collected,
using either a Chain or a Take algorithm (see
section~\ref{sec:basic-oper-lists}), and used as input for subsequent
stages of the reduction, such as IBP reduction, and the decomposition
in terms of known special functions, as described in
section~\ref{sec:reduct-scatt-ampl}.  In our experience, this strategy
is very efficient, even on complex multi-loop integrands, especially
if compared with the more time-consuming IBP reduction step.

It is also worth mentioning that integrand reduction is often used in
combination with \emph{generalized
  unitarity}~\cite{Bern:1994zx,Bern:1994cg,Britto:2004nc,Ellis:2007br,Giele:2008ve}.
On multiple cuts the integrand factorizes as a product of tree-level
amplitudes, which in turn may be evaluated efficiently, over a
numerical field, using Berends-Giele recursion~\cite{Berends:1987me}.
We refer to ref.~\cite{Peraro:2016wsq} for a complete description of
an implementation of generalized unitarity over finite fields.  It
should be noted that, while generalized unitarity is an extremely
powerful method which can substantially reduce the complexity of the
calculation, it also has some limitations.  For instance, one needs to
find rational finite-dimensional parametrizations for the internal
states of the loop on the cut solutions, which is not always easy.
Moreover, in its current state, it cannot be easily applied to
processes with massive internal propagators.  These difficulties and
limitations are not present when applying integrand reduction to a
diagrammatic representation of the amplitude.

\subsection{Choice of an integrand basis}
\label{sec:choice-an-integrand}

It is worth making some observations on possible choices for an
integrand basis.  In the one-loop case, one can choose a basis which
yields a linear combination of known
integrals~\cite{Ossola:2006us,Giele:2008ve}.  With this choice, IBP
reduction is not needed.  At higher loops, this is not the case, and
one should therefore take into account that the elements of an
integrand basis should be later reduced via IBP identities.

A particularly simple but effective choice, especially at the
multi-loop level, consists of writing any on-shell integrand
$\Delta_{\vec{\beta}}$ in terms of the denominators and auxiliaries
$\{D_{T,j}\}$ of its parent topology $T$ such that $\beta_j=0$, i.e.\
excluding the ones that $\Delta_{\vec{\beta}}$ is sitting on.  In
processes with fewer than five external legs, one must also include
scalar products of the form $k_i\cdot \omega_j$ where $\{\omega_j\}$
are a complete set of four-dimensional vectors orthogonal to all the
external momenta $p_1,\ldots,p_e$.  Hence, $\Delta_{\vec{\beta}}$ can
be parametrized as the most general polynomial in this set of
variables, whose total degree is compatible with the theory.  For
instance, a renormalizable theory allows at most one power of loop
momenta per vertex, in the sub-topology defined by the denominators of
$\Delta_{\vec{\beta}}$.  After integrand reduction, the scalar
products of the form $k_i\cdot \omega_j$ can be integrated out in
terms of denominators and auxiliaries $D_{T,j}$.  As explained e.g.\
in~\cite{Mastrolia:2016dhn}, this can be easily done via a tensor
decomposition in the $(d-e+1)$-dimensional subspace orthogonal
to the $e$ external momenta.  Notice that this is very simple even for
complex processes since it only involves the orthogonal projection of
the metric tensor $g_{[d-e+1]}^{\mu \nu}$ and no external momentum.
Alternatively, one can achieve the same result via an angular loop
integration over the orthogonal space, which can be made even simpler
using Gegenbauer polynomials~\cite{Mastrolia:2016dhn}.  This choice of
integrand basis directly yields, after orthogonal integration, a
linear combination of integrals which are suitable for applying
standard IBP identities.  Given also its simplicity, it is a
recommended choice in most cases.

Other choices can be made for the sake of having either a simpler
integrand representation or a larger set of elements of the integrand
basis which integrate to zero.  One can, for instance, choose to
replace monomials in an on-shell integrand with monomials involving
also the extra-dimensional scalar products $\mu_{ij}$.  Because
monomials with $\mu_{ij}$ can be rewritten as linear combinations of
the other ones, one can easily obtain a system of equations relating
these two types of monomials.  By solving this system, assigning
a lower weight to monomials involving $\mu_{ij}$, one can maximize the
presence of integrands which vanish in the four-dimensional limit.
Since we are only interested in a list of independent monomials, it is
sufficient to solve the system numerically (possibly over finite
fields).  This is heuristically found to yield simpler integrand
representations.  However, it also makes IBP reduction harder to use,
since integrands with $\mu_{ij}$ then need to be converted to the ones
in a standard form.  If only the finite part of the amplitude is
needed, one may however choose some integrands involving $\mu_{ij}$
which are $\O(\epsilon)$ after integration and then drop them before
the IBP reduction step.  This may result in notable simplifications.
As an example, at one loop, no on-shell integrand with more than four
denominators contributes to the finite part of an amplitude, if
$\mu_{11}$ is chosen to be the numerator of the five-denominator
integrands in the integrand basis.

Another popular choice is the usage of scalar products involving
momenta which, for an on-shell integrand $\Delta_{\vec{\beta}}$, are
orthogonal to the external momenta of the topology defined by its own
denominators (as opposed to the ones of the parent topology).  One can
indeed build suitable combinations of these scalar products which
vanish upon integration.  Their coefficients can then be dropped after
the integrand reduction.

Another very successful strategy is the usage, for each on-shell
integrand $\Delta_{\vec{\beta}}$, of a complete set of \emph{surface
  terms}, i.e.\ terms which vanish upon integration and are compatible
with multiple
cuts~\cite{Gluza:2010ws,Ita:2015tya,Larsen:2015ped,Abreu:2017hqn}.
These are chosen to be an independent set of IBP equations without
higher powers of denominators.  These define suitable polynomial
numerators for $\Delta_{\vec{\beta}}$ which vanish upon integration.
When this approach is used, IBP reduction is embedded in the integrand
reduction and therefore it is not needed as a separate step.  A
possible disadvantage is that it makes the integrand reduction more
complicated, since these surface terms are typically more complex than
the elements of other integrand bases, and they introduce a dependence
on the dimensional regulator which is otherwise not present in the
integrand reduction stage.  Another disadvantage is that, in the form
it is usually formulated, this strategy can yield incomplete
reductions for some processes.\footnote{Indeed, one can see that, in
  the references above, these surface terms are effectively chosen to
  be linear combinations of IBP identities whose seed integrals do not
  have higher powers of denominators than the ones in the diagrams
  which need to be reduced.  Hence, whenever identities using seed
  integrals with higher powers of denominators, or seed integrals with
  more propagators, are needed to fully reduce a given sector, the
  method above will not yield a complete reduction to a minimal basis
  of master integrals.  Examples where we explicitly checked that
  additional seed integrals are needed are several two-loop topologies
  involving massive internal propagators (e.g.\ topologies for
  amplitudes with two fermion pairs having different masses), and some
  massless four-loop topologies (including most of those reduced in
  ref.~\cite{Henn:2019rmi}).}

We finally point out that, if there is no one-to-one correspondence
between elements of the integrand basis and Feynman integrals to be
reduced via IBPs, one needs to convert between the two.  This step
may also include the transverse integration, if needed.  The
conversion, as in many other cases, can be implemented via a matrix
multiplication.  For this purpose, we recommend using either the Take
And Add algorithm or the Sparse Matrix Multiplication algorithm
described in section~\ref{sec:basic-oper-lists}.

\subsection{Writing the integrand}
\label{sec:writing-integrand}

When using integrand reduction together with Feynman diagrams, one
would typically provide the integrands in Eq.~\eqref{eq:ampint}
analytically.  Even if several methods exist for generating integrands
numerically at one loop, with the notable exception of generalized
unitarity (which is however not based on Feynman diagrams and has the
limitations mentioned above) they have not been generalized to higher
loops.  When integrands are provided in some analytic form, they will
also depend on external polarization vectors, spinor chains, and
possibly other objects describing the external states.  On one hand,
this means that we need to provide a rational parametrization for
these objects.  On the other, we may use the algorithms described
above in order to keep these rational expressions as compact as
possible.  This is done by performing the substitutions which would
yield complex expressions only numerically over finite fields.

A rational parametrization for four-dimensional spinors, polarization
vectors, external momenta, as well as higher-spin polarization states,
can be obtained, in terms of a minimal set of invariants, by means of
the so-called \emph{momentum twistor
  parametrization}~\cite{Hodges:2009hk,Badger:2013gxa,Badger:2016uuq}.
The independent kinematic invariants are called in this case
\emph{momentum twistor variables}.  A comprehensive description of the
usage of this parametrization for describing external states in the
context of numerical calculations over finite fields is given in
ref.~\cite{Peraro:2016wsq} and will not be repeated here.

In amplitudes with only scalars and spin-one external particles, the
only additional loop-dependent objects appearing in the integrand,
besides the loop denominators and auxiliaries, are scalar products
between loop momenta and polarization vectors.  If external fermions
are present, one also has spinor chains involving loop momenta.  These
can be dealt with by splitting the loop momenta in a four-dimensional
and a $(-2 \epsilon)$-dimensional part
\begin{equation}
  \label{eq:20}
  k_j^\mu = k_j^{[4]\, \mu} + k_j^{[-2\epsilon]\, \mu},
\end{equation}
and performing the t'Hooft algebra on the extra-dimensional components
in order to explicitly convert all the dependence on
$k_j^{[-2\epsilon]}$ into the extra-dimensional scalar products
$\mu_{ij}$ defined in Eq.~\eqref{eq:muij}.

The four-dimensional part of the loop momenta is often decomposed into
a four-dimensional basis.  Given a generic loop momentum $k$ and three
massless momenta $p_1,p_2,p_3$, we can use the decomposition, in
spinor notation,
\begin{equation}
  \label{eq:k4ddec}
  k^{[4] \mu} = y_1\, p_1^\mu + y_2\, p_2^\mu + y_3\, \frac{\la 2\, 3\ra}{\la 1\, 3\ra} \frac{\la 1\, \sigma^\mu\, 2 ]}{2} + y_4\, \frac{\la 1\, 3\ra}{\la 2\, 3\ra} \frac{\la 2\, \sigma^\mu\, 1 ]}{2}.
\end{equation}
The massless momenta can be chosen depending on the cut, but it is
also possible, and often easier, to define a global basis of momenta
and therefore use the same set of loop variables $y_j$ and $\mu_{ij}$
everywhere.  If there aren't enough massless external legs, one may
use massless projections of massive ones or arbitrary massless
reference vectors.  In some cases, it is convenient to make the
substitution in Eq.~\eqref{eq:k4ddec} directly in the analytic
integrand, since it provides simplifications for explicit choices of
external helicity states.  In other cases, one may instead make a list
of all the loop-dependent objects (scalar products, spinor chains,
etc\ldots) appearing in the integrand and express them individually as
functions of the variables $y_j$ and $\mu_{ij}$.  This defines a list
of substitutions which can instead be done numerically inside the
linear fit procedure, through the definition of the auxiliary
functions $\mathbf{a}$ appearing in Eq.~\eqref{eq:linfitaux}, while
keeping the integrand written as a rational function of objects which
yield a more compact expression for it.

As we explained, on a multiple cut, the variables $y_j$ and $\mu_{ij}$
are no longer all independent, but a subset of them can be written as
rational functions of the others.  Once again, we note that one does
not need to perform these substitutions explicitly in the integrand,
but only numerically using the auxiliary functions $\mathbf{a}$ as
before.

We finally remark that it is often a good idea to group together
diagrams which share the same denominator structure or can be put
under the same set of denominators as one of the parent topologies of
the process.  Thanks to the fact that, in the linear fit algorithm we
defined in Eq.~\eqref{eq:linfitaux}, we allow an arbitrary sum of
contributions on the r.h.s.,\ this grouping can be easily performed by
including each diagram in this list of contributions (which, we
recall, here includes the integrand and the subtraction terms),
without having to explicitly sum them up analytically.

\section{Decomposition of amplitudes into form factors}
\label{sec:tens-reduct-decomp}

In this section, we briefly discuss the possibility of using the
\textsc{FiniteFlow} framework for an alternative and widely used method for
expressing amplitudes as linear combinations of standard Feynman
integrals.

The method consists of considering an amplitude stripped of all the
external polarization states.  This amplitude will have a set of free
indexes $\lambda_1\ldots,\lambda_e$, which may be Lorentz indexes,
spinor indexes, or other indexes representing higher-spin states.  One
can thus write down the most general linear combination of tensors
$T_j^{\lambda_1\cdots \lambda_e}$ having these indexes, compatible with
the known properties of the amplitude, such as gauge invariance and
other constraints.  More explicitly
\begin{equation}
  \label{eq:Aformfactors}
   A^{\lambda_1\cdots \lambda_e} = \sum_j F_j\, T_j^{\lambda_1\cdots \lambda_e}.
\end{equation}
The \emph{form factors} $F_j$ are rational functions of the kinematic
invariants, which can be computed by contracting the amplitude on the
l.h.s.\ with suitable projectors $P^{\lambda_1\cdots \lambda_e}$
\begin{equation}
  \label{eq:projA}
  F_j = P_j \cdot A,
\end{equation}
where
\begin{equation}
  \label{eq:Tproj}
  P_j^{\lambda_1\cdots \lambda_e} = \sum_k T^{-1}_{jk}\, T_k^{\lambda_1\cdots \lambda_e}
\end{equation}
with
\begin{equation}
  \label{eq:TiTj}
  T_{ij} \equiv T_i\cdot T_j.
\end{equation}
In the previous equations, a dot product between two tensors is a
short-hand for a full contraction between their indexes.

There are at least two bottlenecks in this approach for which the
\textsc{FiniteFlow} framework can be highly beneficial.  The first is the
inversion of the matrix defined in Eq.~\eqref{eq:TiTj}.  This
inversion can be obviously computed using one of the linear solvers
described in section~\ref{sec:dense-sparse-linear} -- typically the
dense solver if the tensors $T_j$ do not have special properties of
orthogonality.  The inversion can also be performed numerically, since
it is only required in an intermediate stage of the calculation, and
can be represented by a node in the dataflow graph.  We find that,
even in cases where the inverse matrix is very complicated, its
numerical inversion takes a negligible amount of time compared with
other parts of the calculation (e.g.\ IBP reduction).  The other
bottleneck which can be significantly mitigated by our framework is
the difficulty of computing the contraction on the r.h.s.\ of
Eq.~\eqref{eq:projA}, in cases where the projectors are particularly
complicated.  Indeed, by substituting Eq.~\eqref{eq:Tproj} into
Eq.~\eqref{eq:projA} we get
\begin{equation}
  \label{eq:projATj}
  F_j = \sum_k T^{-1}_{jk}\, (T_k \cdot A).
\end{equation}
This means that we can compute the contractions $T_k \cdot A$ instead,
which are usually significantly simpler, and multiply them
(numerically) by the matrix $T^{-1}_{jk}$ at a later stage.  This
allows to reconstruct the form factors directly without ever needing
explicit analytic expressions for the projectors.  One can further
elaborate the algorithm by contracting the free indexes of
Eq.~\eqref{eq:Aformfactors} with explicit polarization states, for the
direct reconstruction of helicity amplitudes rather than the form
factors themselves.

\section{Finding integrable symbols from a known alphabet}
\label{sec:find-integr-symb}

As we already stated, many Feynman integrals can be cast as iterated
integrals in the form of Eq.~\eqref{eq:dlogiterated}.  It is customary
to associate to these integrals an object called
\emph{symbol}~\cite{Goncharov:2010jf,Duhr:2011zq}.  For the purposes
of this paper, we define the symbol as
\begin{equation}
  \label{eq:symbol}
  \mathcal{S}\left( \int d \log w_1 \circ d \log w_2 \circ \cdots  \right) \equiv w_1 \otimes w_2 \otimes \cdots,
\end{equation}
where, as already mentioned in section~\ref{sec:diff-equat-epsil},
$w_k$ are called \emph{letters}, and a complete set of letters
$W=\{w_k\}$ is called \emph{alphabet}.  Because the symbol does not
depend on the integration path and the boundary terms, it contains
less information than the full iterated integral, but it is still a
very interesting object to study for determining the analytic
structure of an amplitude.  More information on symbols, their
properties, and their relations to multiple polylogarithms can be
found in~\cite{Duhr:2011zq}.

Given a known alphabet $W$, one can build symbols of \emph{weight} $n$
as linear combinations of those defined in Eq.~\eqref{eq:symbol},
namely
\begin{equation}
  \label{eq:symbollc}
  S = \sum_{j_1,\ldots, j_n} c_{j_1\cdots j_n}\, w_{j_1} \otimes \cdots \otimes w_{j_n}.
\end{equation}
However, in general, such a linear combination is not
\emph{integrable}, i.e.\ it does not integrate to a function which is
independent of the integration path.  As pointed out
in~\cite{Brown:2009qja}, a necessary and sufficient condition for the
symbol in Eq.~\eqref{eq:symbollc} to be integrable is
\begin{equation}
  \label{eq:integrcond}
  \sum_{j_1,\ldots, j_n} c_{j_1\cdots j_n} \left(  \frac{\partial \log w_{j_k}}{\partial z_l}\frac{\partial \log w_{j_{k+1}}}{\partial z_m} - \left( l \leftrightarrow m \right) \right)\, w_{j_1} \otimes \cdots \otimes \hat w_{j_k} \otimes \hat w_{j_{k+1}} \otimes \cdots \otimes w_{j_n}=0,
\end{equation}
for all $k=1,\ldots,n-1$ and all pairs $(z_l,z_m)$, where
$\z = \{z_j\}$ are the kinematic variables the letters depend on.  In
the previous equation, $\hat w_k$ indicates the omission of the letter
$w_k$.  By solving these integrability conditions, which amounts to
solve a linear system for the coefficients $c_{j_1\cdots j_n}$, one
can build a complete list of integrable symbols of weight $n$.  It is
worth mentioning that there are additional conditions one can impose
to restrict the number of terms in the ansatz of
Eq.~\eqref{eq:symbollc}, namely additional conditions on the allowed
entries of a symbol.  For instance, the first entry, which is related
to the discontinuity of the function, may be restricted to contain
only letters associated to physical branch points of an amplitude.

Here we discuss a simple method\footnote{This method has been
  independently developed by the author and used in several
  unpublished tests and checks~(see e.g.\ ref.~\cite{Zoia:2018sin}).
  It shares some similarities with the one implemented
  in~\cite{Mitev:2018kie}.}  for finding all integrable symbols from a
known alphabet, up to a specified weight $n$, exploiting the
algorithms of the framework we presented in this paper.

We first observe that the only dependence of Eq.~\eqref{eq:integrcond}
on the explicit analytic expressions of the letters is via the crossed
derivatives
\begin{equation}
  \label{eq:14}
  d^{(lm)}_{ij} \equiv \frac{\partial \log w_{i}}{\partial z_l}\frac{\partial \log w_{j}}{\partial z_m} - \left( l \leftrightarrow m \right).
\end{equation}
In order to simplify the notation, let us define a multi-index
$J=(i,j,l,m)$ such that
\begin{equation}
  \label{eq:17}
  d_J \equiv d^{(lm)}_{ij}.
\end{equation}
The only relevant information about these derivatives which is needed
for the purpose of solving Eq.~\eqref{eq:integrcond} are possible
linear relations which may exist between different elements $d_J$.
These relations only depend on the alphabet, and not on the weight of
the symbols which need to be considered.  Once all these linear
relations have been found for a given alphabet, the integrability
conditions can be solved at any weight using a numeric linear system
over $\Q$, and without using the analytic expressions of the letters
again.

In order to find these linear relations, we first compute analytic
expressions for all the functions $d_J$, which can usually be done in
seconds even for complex alphabets.  If the functions $d_J$ have no
square root in them, we simply solve the \emph{linear-fit} problem
\begin{equation}
  \label{eq:linindepdj}
  \sum_{J} x_J d_J = 0,
\end{equation}
where the unknowns $x_J$ are $\Q$-numbers, while the functions $d_J$
depend on the variables $\z$.  This equation is solved with respect to
the unknowns $x_J$ using the (numerical version of the) linear fit
algorithm already described in this paper.  Linear relations between
the unknowns $x_J$ are thus easily translated into relations between
the functions $d_J$ (notice that independent unknowns multiply
dependent functions, and the other way around).  In order to simplify
the linear fit, it is convenient to extract a priori some obvious
relations, such as relations of the form $d_J=0$ or
$d_{J_1}=\pm d_{J_2}$, which are more easily identifiable from the
analytic expressions.

If the functions $d_J$ depend on a set of (independent) square roots,
we first rewrite each of them in a canonical form, such that each
function is multi-linear in the square roots.  This can be easily
done, one square root at the time, by replacing a given square root
$\sqrt{f}$ with an auxiliary variable, say $r$, and computing the
remainder of $d_J=d_J(r)$ with respect to $r^2-f$, via a univariate
polynomial division with respect to $r$ (note that univariate
polynomial remainders are easily generalized to apply to rational
functions\footnote{For instance, one can use the built-in
  \textsc{Mathematica} procedure \texttt{PolynomialRemainder}, which
  applies to rational functions as well.}).  The result will be linear
in $r$.  If all the square roots are chosen to be independent, after
putting the functions $d_J$ in this canonical form, one can simply
solve the linear fit in Eq.~\eqref{eq:linindepdj} by replacing each
square root with a new independent variable.  This works because, if
Eq.~\eqref{eq:linindepdj} holds and the $d_J$ are put in this
canonical form, then the terms multiplying independent monomials in
the square roots must vanish separately.  This is effectively
equivalent to performing a linear fit where each square root is
treated as an independent variable.  We find that, even in cases where
square roots are rationalizable, this approach is often more efficient
than using a change of variables which rationalizes the square roots.

Once a complete set of linear relations between the crossed
derivatives $d_J$ has been found, one can use this information alone
to solve the integrability conditions.  This is best done recursively
from lower to higher weights.  As already stated, we understand that
other conditions may still restrict the ansatz at any weight and
therefore the list of integrable symbols.

At weight $n=1$, every letter trivially defines an integrable symbol.
At higher weights, it is customary to exploit the lower weight
information in order to build a smaller ansatz than the one in
Eq.~\eqref{eq:symbollc}.  If $\{S_j^{(n-1)}\}_j$ is a complete set of
integrable symbols at weight $n-1$, we find the integrable symbols
$S_j^{(n)}$ at weight $n$ as follows.  We write our ansatz as
\begin{equation}
  \label{eq:intsymansatz}
  S = \sum_{jk}\, c_{jk}\, S_j^{(n-1)}\otimes w_k.
\end{equation}
Because the symbols $S_j^{(n-1)}$ are already integrable, we only need
to impose the integrability condition on the last two entries.  Hence,
for all possible pairs of variables $(z_l,z_m)$ we make the
substitution
\begin{equation}
  \label{eq:intcondsubst}
  w_{j_1} \otimes \cdots \otimes w_{j_n} \to (w_{j_1} \otimes \cdots \otimes w_{j_{n-2}})
  \, d_{j_{n-1} j_n}^{(lm)}
\end{equation}
into Eq.~\eqref{eq:intsymansatz}, while $d_{j_{n-1} j_n}^{(lm)}$ are
left as arbitrary variables (i.e.\ without explicitly substituting
their expressions, which are no longer relevant at this stage).  Then
we substitute the linear relations satisfied by the
$d_{j_{n-1} j_n}^{(lm)}$ such that our ansatz is written in terms of
linearly independent functions (still represented by independent
variables in the formulas), and we impose that the coefficient of each
independent structure with the form of the r.h.s.\
of~\eqref{eq:intcondsubst} vanishes.  This strategy builds a numeric
\emph{sparse linear system} of equations for the coefficients $c_{jk}$
in Eq.~\eqref{eq:intsymansatz}, which can be solved with the algorithm
already discussed in this paper.  Linear relations between the
coefficients $c_{jk}$ are then easily translated into a set of
linearly independent symbols at weight $n$ satisfying the
integrability conditions.

\section{Proof-of-concept implementation}
\label{sec:proof-conc-impl}

With this paper, we also publicly release a proof-of-concept
implementation of the \textsc{FiniteFlow} framework.  The code is available
here
\begin{equation*}
  \textrm{\url{https://github.com/peraro/finiteflow}}
\end{equation*}
and can be installed and used following the instructions given at that
URL.  This code is the result of experimentation and trial and error,
and should not be regarded as an example of high coding standards or
as a final implementation of this framework.  Despite this, it has
already been used for obtaining several cutting-edge research results
in high energy physics, and we believe its public release can be
highly beneficial to the community.  It also includes the
\textsc{FiniteFlow} package for \textsc{Mathematica}, which provides
a high-level interface to the routines of the library.

We also release a collection of packages and examples using the
\textsc{Mathematica} interface to this code, at the URL
\begin{equation*}
  \textrm{\url{https://github.com/peraro/finiteflow-mathtools}}
\end{equation*}
which includes several applications described in this paper.  In
particular, it contains the following packages:
\begin{description}
\item[FFUtils] Utilities implementing simple general purpose
  algorithms, such as algorithms for finding linear relations between
  functions.
\item[LiteMomentum] Utilities for momenta in Quantum Field Theory.  It
  does not use \textsc{FiniteFlow}, but it is used by other packages
  and examples in the same repository.
\item[LiteIBP] Utilities and tools for generating IBP systems of
  equations and differential equations for Feynman integrals, to be
  used together with the \textsc{LiteRed}~\cite{Lee:2012cn} package.
\item[Symbols] Scripts for building integrable symbols from known
  alphabets.
\end{description}
We note that these packages should be regarded as a set of utilities
rather the implementation of fully automated solutions for specific
tasks.  They are also meant as examples of how to build packages on
top of the \textsc{Mathematica} interface to the code.  The same
repository also contains several examples of usage of the
\textsc{FiniteFlow} package.  While these examples have been chosen to
be simple enough to run in a few minutes on a modern laptop, they can
be used as templates to be adapted to significantly more complex
problems.  We therefore recommend reading the documentation which
comes with them and the comments inside their source as an
introduction to the usage of this code for the applications described
in this paper.

In this section, we give some details on some aspects and features of
our implementation of the \textsc{FiniteFlow} framework and provide
some observations about possible improvements for the future.

The code is implemented in \textsc{C++} and we provide a high-level
\textsc{Mathematica} interface.  At the time of writing, the
\textsc{Mathematica} interface is the easiest and more flexible way of
using \textsc{FiniteFlow}, since it allows to combine the features of our
framework with the ones of a full computer algebra system.  Interfaces
to other high-level languages, such as \textsc{Python}, and computer
algebra systems are likely to be added in the future.

This implementation uses finite fields $\Z_p$ where $p$ are 63-bit
integers.  We have explicitly hard-coded a list of primes satisfying
$2^{63}>p>2^{62}$ -- namely the 201 largest primes with this property
-- which define all the finite fields we use.  In particular, by
making the assumption that all the primes we use belong to that range,
we are able to perform a few optimizations in basic arithmetic
operations.  We use a few routines and macros of the \textsc{Flint}
library for basic operations of modular arithmetic (we also optionally
provide a heavily stripped down version of \textsc{Flint} with only
the parts which are needed for \textsc{FiniteFlow}, with fewer
dependencies, as well as an easier and faster installation), such
as the calculation of multiplicative inverses and modular
multiplication using extended precision and precomputed
reciprocals~\cite{ImprovedDivision}.

We use several representations of polynomials and rational functions,
depending on the task.  As already explained in
section~\ref{sec:eval-polyn-rati}, if we need to repeatedly evaluate
polynomials and rational functions, we store the data representing
them as a contiguous array of integers and evaluate them by means of
the Horner scheme.  For polynomials in Newton's representation, we
store an array with the sequence $\{y_j\}$ and another one with the
coefficients $a_j$.  The latter is an array of integers in the
univariate case (see Eq.~\eqref{eq:unewtonrep}) and an array of Newton
polynomials in fewer variables in the multivariate case (see
Eq.~\eqref{eq:mnewtonrep}).  Univariate rational functions in Thiele's
representation (given in Eq.~\eqref{eq:thielerep}) are stored similarly
to univariate Newton polynomials.  For every other task, we use a
sparse polynomial representation which consists of a list of
non-vanishing monomials.  Each monomial is, in turn, a numerical
coefficient (in $\Z_p$ or $\Q$) and an associated list of exponents
for the variables.  This representation is used for most algebraic
operations on polynomials, e.g.\ when converting Newton's polynomials
in a canonical form, or when shifting variables (we recall that a
shift of variables is typically required by the functional
reconstruction algorithm we use).  It is also the most convenient
representation for communicating polynomial expressions between
\textsc{FiniteFlow} and other programs such as computer algebra systems.

Our system for dataflow graphs distinguishes several types of objects,
namely \emph{sessions}, \emph{graphs}, \emph{nodes} and
\emph{algorithms}.

Sessions are objects which contain a list of graphs and are
responsible for doing most operations using them, such as evaluating
them while handling parallelization, and running functional
reconstruction algorithms.  Since a session can contain any number of
dataflow graphs, for most applications there is no reason for using
more than one session in the same program, although it is obviously
possible.  The concept of a session is not (explicitly) present in the
\textsc{Mathematica} interface since the latter only uses one global
session.  Graphs in the same session, as well as nodes in a graph, are
associated with a non-negative integer ID.  In the \textsc{Mathematica}
interface, these IDs can instead be any expression, which is seamlessly
mapped to the correct integer ID when communicating with the
\textsc{C++} code.  Graphs, as already explained, are collections of
nodes.  Nodes are implemented as wrappers around algorithms and
contain a list of IDs corresponding to their inputs.  When building a
new node for a graph, the program checks that the expected lengths of
its input lists are consistent with the ones of the output lists of
its input nodes.  Algorithms are the lowest-level objects responsible
for the numerical evaluations, and they have associated procedures for
it.  Algorithms might also have a procedure for their learning phase
and, in that case, they also specify how many times this should be
called (with different inputs).

Because an algorithm might have to run in parallel for different input
values, it is made of two types of data.  The first type is read-only
data, i.e.\ data which is not specific to an evaluation point and can
be shared across several threads during parallelization.  This might
also include data which is mutable only during the learning phase.
The second type of data can instead be modified during any numerical
evaluation.  In multi-threaded applications, mutable data needs to be
cloned across all the threads in order to avoid data races.  Algorithm
objects thus have associated routines for cloning mutable data.

In the future, we might further split mutable data into two types.
The first is mutable data which only depends on the finite field
$\Z_p$.  This data only needs to be copied a number of times equal to
the maximum number of fields used at the same time in a parallel
evaluation, which is typically no larger than two.  The second is data
which can depend on both the prime $p$ and the variables $\z$ which
are the input of a given graph.  Only for the latter one needs to make
a copy for each thread.  Therefore, even though it is not currently
implemented, this further split can improve memory usage by
significantly reducing the amount of cloned data.  As an example,
consider a linear system with parametric entries depending on
variables $\z$.  The rational functions defining the entries of the
system as rational functions over $\Q$, as well as the list of
independent unknowns and equations, are immutable data.  The same
functions mapped with over $\Z_p$ depend on the prime $p$ but not on
the points $\z$.  Finally, the numerical system, obtained by
evaluating such functions numerically for specific inputs $\z$,
depends on both the prime field and the evaluation point.

We point out that the usage of dataflow graphs also greatly simplifies
multi-threading.  It is indeed sufficient that each type of basic
algorithm has an associated procedure for cloning its non-mutable
data.  From these, the framework is able to automatically clone the
mutable data of any complex graph, and correctly use it for the
purpose of performing multi-threaded evaluations.  A similar potential
advantage regards serialization of algorithms, although this feature
is not implemented at the time of writing.  In principle, each basic
algorithm may have an associated procedure for serializing and
deserializing its data.  From these, one would be able to serialize
complete graphs representing arbitrarily complex calculations.  This
could be useful for both sharing graphs and loading them up more
quickly, together with the information about the learning phases which
have already been completed.

We now turn to the caching system used to store the evaluations of a
graph.  We recall that, in the multivariate case, we start by
performing some preliminary univariate reconstructions, which
determine (among other things) a list of evaluation points needed to
reconstruct the output of a graph.  In principle, for each evaluation
point, we may need to store the input variables, the whole output list
of the graph, and the prime $p$ which defines the finite field.
Unfortunately, when the output of a graph is a long list and a large
number of evaluation points is needed, this straightforward strategy
can yield issues related to memory usage.  This can be true even when
a Non-Zeros node is appended to a graph~(see
section~\ref{sec:basic-oper-lists}), as we have already recommended.
Hence, we adopt a slightly more refined strategy which works well in
realistic scenarios.  Heuristically, we observe that, when the output
of a graph is a long list, the complexity of the elements of the list
can vary significantly.  In particular, many elements correspond to
relatively simple rational functions while, usually, only a few of
them have high complexity.  Simpler rational functions obviously need
fewer evaluation points in order to be reconstructed.  Hence, one
could improve this strategy by storing a shorter output list
containing, for each evaluation point, only the elements of the output
which need that point for their reconstruction.  In practice, we
proceed as follows.  Once a complete list of evaluation points has
been determined, for each element of the output we tag all the points
needed for its reconstruction.  In our implementation, this tagging
requires one bit of memory for each output element.  If an evaluation
point is never tagged, it is removed from the list.  Then, after a
graph is evaluated on a given point, we only store the entries of the
output for which that point is needed.  This typically allows to store
a much shorter output list on most evaluation points, therefore
yielding a major improvement in memory usage.  When combined with
Non-Zeros nodes, we find that with this strategy the caching of the
evaluations is hardly ever a bottleneck in terms of memory usage,
especially when the code is run on high-memory machines available in
clusters and other computing facilities often used for intensive
scientific computations.

We also point out that, as explained more in detail in
section~\ref{sec:parallel-execution}, one can generate lists of needed
evaluation points and separately evaluate subsets of them, either
sequentially or in parallel.  On top of being a powerful option for
parallelization, this feature also allows to split long calculations
into smaller batches and save intermediate results to disk, such that
they are not lost in case of system crashes or other errors which may
prevent the evaluations to successfully complete.

The \textsc{FiniteFlow} library implements the basic numerical algorithms
described in this paper, the functional reconstruction methods we
discussed, as well as the framework based on dataflow graphs.  When
the latter is used, one can easily define complex numerical algorithms
without any low-level coding.  This can be done even more easily from
the \textsc{Mathematica} interface.  The latter also offers some
convenient wrappers for common tasks, such as solving analytic or
numeric linear systems or linear fits.  These wrappers hide the
dataflow-based implementation.  However, as discussed in this paper,
the approach based on dataflow graphs offers a flexibility which
greatly enhances the scope of possible applications of this framework.

The approach based on dataflow graphs is the preferred way of defining
algorithms with the library, especially when using the
\textsc{Mathematica} interface.  However, the library can also be
enhanced by custom numerical algorithms written in \textsc{C++}.  For
instance, the results presented
in~\cite{Badger:2017jhb,Badger:2018enw} used a custom \textsc{C++}
extension of the linear fit algorithm which computes generalized
unitarity cuts via Berends-Giele currents, as explained in
ref.~\cite{Peraro:2016wsq} (this extension is not included in the
public code).

It should also be clear that the \textsc{FiniteFlow} framework is not
designed to solve one specific problem, but as a method to implement
solutions for a large variety of algebraic problems.  By building on
top of this public code, one can, of course, implement higher-level and
easier-to-use solutions for more specific tasks.

\subsection{Parallel execution}
\label{sec:parallel-execution}

As discussed in section~\ref{sec:parallelization}, one of the main
advantages of functional reconstruction algorithms is that they can be
massively parallelized.  In our current implementation, we offer two
strategies for parallelization, which can also be used together.

The first and easier-to-use strategy is \emph{multi-threading}.  This
is handled completely automatically by the code when the
dataflow-based approach is used.  Data which cannot be shared among
threads is cloned as needed and parallelization is achieved by
splitting the calculation over an appropriate number of threads, as
explained in section~\ref{sec:parallelization}.  The number of threads
which is used can either be specified manually or chosen automatically
based on the hardware configuration.  We recommend specifying it
manually when using the code on clusters or machines shared among
several users, since the automatic choice might not be the most
appropriate one in such cases.

The second method allows to further enhance parallelization
possibilities by using several nodes of a cluster, or even several
(possibly unrelated) machines, for the evaluations of the function to
be reconstructed.  In order to use this method, after defining a
numerical algorithm, we compute and store the total and partial
degrees of its output.  As explained, this is done via univariate
reconstructions which are much quicker than a full multivariate one.
From this information, we also build and store a list of inputs for
the evaluations.  For this, we need to make a guess of how many prime
fields will be needed.  One can, however, start by assuming only one
prime field is needed, and add more points at a later time if this is
not the case.  The stored list of needed evaluation points can be
shared across several nodes or several machines, where any subset of
them can be computed and saved independently.  Of course, these
evaluations can (and will, by default) be further parallelized using
multi-threading, as discussed above.  Finally, the evaluations are
collected on one machine where the reconstruction is performed.
Should the reconstruction fail due to the need of more prime fields,
we increase our guess on the number of primes needed and create a
complementary list of evaluation points.  We then proceed with the
evaluation of these additional points, across several nodes or
machines as for the previous one, and collect them for the
reconstruction.  We proceed this way until the reconstruction is
successful.  This method greatly increases the potential
parallelization options, at the price of being less automated, since
the lists of evaluations need to be generated and copied around by
hand.\footnote{ In the future, we might consider implementing other
  approaches, such as the use of the standard \textsc{Message Passing
    Interface} (\textsc{MPI}) to offer a more automated way of
  parallelizing the evaluations across several nodes of the same
  cluster.  However, the latter approach would end up being more
  limiting than the one we already implemented, since \textsc{MPI}
  does not support parallelization over several unrelated machines.}
This option can be very beneficial for reconstructing particularly
complex functions, or functions whose numerical evaluation is very
time-consuming.  As already mentioned, it also provides a method
splitting up long calculations in smaller batches and saving
intermediate results on disk.

\section{Conclusions}
\label{sec:conclusions}

We presented the \textsc{FiniteFlow} framework, which establishes a novel
and effective way of defining and implementing complex algebraic
calculations.  The framework comprises an efficient low-level
implementation of basic numerical algorithms over finite fields, a
system for easily combining these basic algorithms into computational
graphs -- known as dataflow graphs -- representing arbitrarily complex
algorithms, and multivariate functional reconstruction techniques for
obtaining analytic results of out these numerical evaluations.

Within this framework, complex calculations can be easily implemented
using high-level languages and computer algebra systems, without being
concerned with the low-level details of the implementation.  It also
offers a highly automated way of parallelizing the calculation, thus
fully exploiting available computing resources.

The framework is easy to use, efficient, and extremely flexible.  It
can be employed for the solution of a huge variety of algebraic
problems, in several fields.  It allows to directly reconstruct
analytic expressions for the final results of algebraic calculations,
thus sidestepping the appearance of large intermediate expressions,
which are typically a major bottleneck.

In this paper, we have shown several applications of this framework
to highly relevant problems in high-energy physics, in particular
concerning the calculation of multi-loop scattering amplitude.

We also release a proof-of-concept implementation of this framework.
This implementation has already been successfully applied to several
state-of-the-art problems, some of which proved to be beyond the reach
of traditional computer algebra, using reasonable computing resources.
Notable examples are recent results for two-loop five-gluon helicity
amplitudes in Yang-Mills theory~\cite{Badger:2018enw,Badger:2019djh}
and the reduction of four-loop form factors to master
integrals~\cite{Henn:2019rmi}.  We point out that these two types of
examples are complex for very different reasons.  In the former, a
large part of the complexity is due to the high number of scales,
while in the latter, which only has one scale, it is due to the huge
size of the IBP systems one needs to solve.  Quite remarkably, the
techniques described in this paper have been able to tackle both these
cases, showing that they are capable of dealing with a wide spectrum
of complex problems.

We believe the algorithms presented in this paper, and their publicly
released proof-of-concept implementation, will contribute to pushing
the limits of what is possible in terms of algebraic calculations.
Due to their efficiency and flexibility, they will be useful in the
future for obtaining more scientific results concerning a wide range
of problems.




\section*{Acknowledgements}
I thank Simon Badger, Johannes Henn, Pierpaolo Mastrolia, and Lorenzo
Tancredi for many discussions, comments, and for their collaboration
on topics which motivated the development of the methods presented in
this paper.  I am also grateful to Simon Badger, Christian
Brønnum-Hansen, Heribertus Bayu Hartanto, and William Torres
Bobadilla for testing various features of the proof-of-concept
implementation of these methods and providing valuable feedback.  This
project has received funding from the European Union’s Horizon 2020
research and innovation programme under the Marie Skłodowska-Curie
grant agreement 746223.

\begin{appendices}

\section{Mutability of graphs and nodes}
\label{sec:mutab-graphs-nodes}

In this appendix we discuss a technical aspect of our implementation
of dataflow graphs, namely the \emph{mutability} of graphs and nodes.
In general, mutating nodes or graphs which have already been defined
can lead to inconsistencies between the implemented algorithms and
their expected inputs.  However, depending on the use case, it may be
convenient to have the ability of performing such mutations, to the
extent that the defined graphs are always consistent.

Let us first discuss the mutability of \emph{nodes}.  Mutating a node
can mean either deleting it, replacing it, or modifying its metadata
in a way that changes its output (e.g.\ changing the list of needed
unknowns in a linear system).  We find it is convenient to allow such
mutations, as long as a node is not used as input in any other node of
a graph.  Once a node $N_1$ is specified as input of another node
$N_2$, the input node $N_1$ becomes \emph{immutable}, i.e.\ the
mutations described above are no longer allowed.  This is done in
order to prevent changes of the length of the output of node $N_1$
which can make the evaluation of node $N_2$ impossible (note that in
principle we may allow swapping two nodes which have the same lengths
for the input and output lists, but this is currently not
implemented).  As a conveniency, we allow to make node $N_1$ mutable
again, after node $N_2$ and all other nodes using $N_1$ as input have
been deleted.

We now turn to the mutability of \emph{graphs}.  In particular, this
is relevant when using \emph{subgraph} nodes.  Mutating a graph may
involve adding, deleting, or mutating its nodes, and changing its
output node.  Once a graph $G_1$ is specified as subgraph of a node
$N$ in another graph $G_2$, then the graph $G_1$ becomes immutable.
If this was not the case, mutations to the graph $G_1$ may modify its
output and make node $N$, and therefore graph $G_2$, impossible to
evaluate.  For the same reason, the output node of $G_1$ is also made
immutable in such cases.  Once all the nodes using $G_1$ as subgraph
are deleted, graph $G_1$ becomes mutable again.

\section{Further observations on IBP identities}
\label{sec:observ-integr-parts}

In this appendix we collect some observations about IBP identities
which complement the discussion in
section~\ref{sec:reduct-mast-integr}.

We already observed that, in order to solve any linear system, we must
sort the unknowns by weight.  Whenever we solve an equation, higher
weight unknowns are expressed in terms of lower weight unknowns.  The
complexity of the Gauss elimination algorithm for a sparse system can
strongly depend on this choice of weight.  Therefore, even if any
choice of weight can be specified when defining a system, it is
worth giving an example which we found works well for IBP systems.

In the case of an IBP system, the unknowns are Feynman integrals.  For
the purpose of assigning a weight to them, it is customary to
associate to each integral in Eq.~\eqref{eq:canonicalint} the
following numbers:
\begin{itemize}
\item $t$ is the number of exponents $\alpha_j$ such that $\alpha_j>0$
\item $r$ is the sum of the positive exponents
  \begin{equation}
    \label{eq:4}
    r = \sum_{j | \alpha_j > 0} \alpha_j
  \end{equation}
\item $s$ is minus the sum of the negative exponents
    \begin{equation}
    s = -\sum_{j | \alpha_j < 0} \alpha_j \geq 0.
  \end{equation}
\end{itemize}
It is generally understood that the higher these numbers are, the more
complex an integral should be considered~\cite{Laporta:2001dd}.  It is
also customary to use the notion of \emph{sector} of an integral,
which is identified by the list of indexes $j$ such that the exponents
$\alpha_j$ are positive, i.e.\ $\{j | \alpha_j>0\}$.  In other words,
two integrals belonging to the same sector depend on the same list of
denominators, possibly raised to different powers, and possibly with a
different numerator.  As an example, a definition of weight for
Feynman integral can be determined, by the following criteria, in
order of importance:
\begin{itemize}
\item the positive integer $r-t$, where a higher number means higher weight
\item the positive integer $t$, where a higher number means higher weight
\item the positive integer $r$, where a higher number means higher weight
\item the positive integer $s$, where a higher number means higher weight
\item integrals in a topology $T_1$ are considered to be of higher
  weight if they belong to a sector mapped to a different topology
  $T_2$
\item integrals in a sector of a topology $T$ are considered to be of
  higher weight if they belong to a sector mapped to another sector of
  the same topology
\item the positive integer $\textrm{max}(\{-\alpha_j\}_{j | \alpha_j < 0})$,
  where a higher integer means higher weight.
\end{itemize}
If the criteria above are not sufficient to uniquely sort two
different integrals, we fall back to any other criterion which defines
a total ordering, such as the intrinsic ordering built in a computer
algebra system to sort expressions.  The choice above prefers
integrals with powers of denominators no higher than one -- indeed,
this is used as the very first criterion for determining the weight of
a Feynman integrals.  We found that this choice is particularly
effective when combined with the mark-and-sweep algorithm for
filtering out unneeded equations, since it often yields a smaller set
of needed equations than other choices.  We however stress again that,
of course, many other definitions of weight are possible and can be
specified instead of the one suggested here.

We make a few more observations about the generation of IBP systems.
These equations -- which include IBPs, Lorentz invariance identities,
symmetries among integrals of the same sectors, and mappings between
integrals of different sectors -- are typically first generated for
generic Feynman integrals of the form of Eq.~\eqref{eq:canonicalint}
with arbitrary symbolic exponents.  These are sometimes called
\emph{template equations}.  The IBP system is thus generated by
writing down these template equations for specific Feynman integrals
(i.e.\ for specific values of the exponents), which in this context
are known as \emph{seed integrals}.  It is interesting to understand
how many and which seed integrals must be chosen in order to
successfully reduce a given set of needed integrals to master
integrals.  To the best of our knowledge, there is no way of
determining a priori a minimal choice which works, but common choices
which are expected to work in most cases (despite not being minimal)
exist.  A popular choice which usually works is selecting a range for
the integers $s$ and $r$ of the seed integrals, based on the choice
one must make for the top-level sectors.  However, we find that it is
often more convenient to specify a range in $s$ and $r-t$ instead.  In
particular, for most topologies one only needs to select seed
integrals for which the value of $r-t$ is either the same or one unity
higher that the maximal one between the integrals which need to be
reduced.  We however also point out that, while an over-conservative
choice of seed integrals will result in a slowdown of the learning
phase, the equations generated from unneeded seed integrals may be all
successfully filtered out by the mark-and-sweep algorithm, hence
reducing the system to the same one would have obtained with a more
optimal choice.  However, we also point out that this may or may not
happen depending on the chosen ordering for the Feynman integrals.  We
have empirically observed that it does happen for the choice of
ordering based on the definition of weight we suggested above.

We conclude this appendix with an observation about sector mappings
which we haven't found elsewhere in the literature.  This concerns
kinematic configurations which have symmetries with respect to
permutations of external legs, i.e.\ permutations of external momenta
which preserve all the kinematic invariants.  Notable examples are
three-point kinematics with two massless legs, and four-point fully
massless kinematics.  For these kinematic configurations we can
distinguish two types of sector mappings.  The first one, which we
call here \emph{normal mappings}, simply consists of shifts of the loop
momenta which map a sector into a different one.  The second one,
which we call \emph{generalized mappings}, consists of a permutation of
external legs which preserves the kinematic invariants, optionally
followed by a shift of the loop momenta.  The most typical approach to
deal with these mappings does not distinguish between the two types.
In particular, for all mapped sectors, only sector mappings are
generated in the system of equations, and no IBP identity, Lorentz
identity or sector symmetry.  The rationale is that one would expect
the other identities to be automatically covered by combining sector
mappings with identities generated for the unique (unmapped) sectors.
However, we explicitly verified that this is not always the case for
generalized mappings.  In other words, given a set of seed integrals
for a generalized mapped sector, there are some identities which are
independent of the ones generated by combining sector mappings for the
same set of seed integrals, and identities for the unique sectors.
The missing identities can be recovered by adding more seed integrals
to the mapped sectors and to the unique sectors, at the price of
obtaining a more complex system of equations.  Notice that this is
similar to what happens for Lorentz invariance identities, which in
principle can be replaced by IBP identities only, at the price of
using more seed integrals and making the system more complex.  A
simple example of this is the two-loop massless double box.  We indeed
found that this topology can be reduced to master integrals, for any
range in $s$, by considering only seed integrals with $r-t=0$, as long
as IBPs and Lorentz invariance identities are generated also for
sectors satisfying generalized mappings.  When these additional
identities are not included, we need to add seed integrals with
$r-t=1$ in order to successfully perform the reduction.  We therefore
recommend to generate, alongside generalized mappings, also IBPs,
Lorentz identities and symmetries for sectors which satisfy them.
This is even more convenient when using the mark-and-sweep algorithm
for simplifying the system, since the simpler equations with lower
$r-t$ are automatically selected if available.  This can eventually
yield a smaller system with easier equations to solve.  For similar
reasons, we recommend to always add Lorentz invariance identities,
regardless of the topology.

\end{appendices}

\bibliographystyle{JHEP}
\bibliography{biblio}

\end{document}